\documentclass{article}
\usepackage{amsfonts}

\input{tcilatex}
\begin{document}

\begin{center}
\bigskip \textbf{Veli Shakhmurov}

Department of Mechanical Engineering, Okan University, Akfirat, Tuzla 34959
Istanbul, Turkey,

E-mail: veli.sahmurov@okan.edu.tr\ 

\textbf{On the dynamics of a cancer tumor growth model with multiphase
structure}

\bigskip

\textbf{Abstract}
\end{center}

In this paper, we study a phase-space analysis of a mathematical model of
tumor growth with an immune response. Mathematical analysis of the model
equations with multipoint initial condition, regarding to dissipativity,
boundedness of solutions, invariance of non-negativity, nature of
equilibria, local and global stability will be investigated. We study some
features of behavior of one three-dimensional tumor growth model with
dynamics described in terms of densities of three cells populations: tumor
cells, healthy host cells and effector immune cells. We find the upper and
lower bounds for the effector immune cells population, with $t\rightarrow
\infty $. Further, we derive sufficient conditions under which trajectories
from the positive domain of feasible multipoint initial conditions tend to
one of equilibrium points. Here cases of the small tumor mass equilibrium
point; the healthy equilibrium point; the \textquotedblleft
death\textquotedblright\ equilibrium point are examined. Biological
implications of our results are considered.


\begin{center}
\textbf{1. Introduction}
\end{center}

Beginning with this article we intend to attempt to investigate the problems
of Mathematical and Biological approaches to modelings of cancer growth
dynamics processes and operations. It is important to take into account
\textquotedblleft the nonlinear property of cancer growth
processes\textquotedblright\ in construction of mathematical logistic
models. This nonlinearity approach appears very convenient to display
unexpected dynamics in cancer growth processes, expressed in different
reactions of the dynamics to different concentrations of immune cells at
different stages of cancer growth developments $\left[ 1-10\right] $. Taking
into account all the complex processes, nonlinear mathematical models can be
estimated capable of compensation and minimization the inconsistencies
between different mathematical models related to cancer growth-anticancer
factor affections. The elaboration of mathematical non-spatial models of the
cancer tumor growth in the broad framework of tumor immune interactions
studies is one of intensively developing areas in the modern mathematical
biology, see works $[1-7]$. Of course, the development of powerful cancer
immunotherapies requires first an understanding of the mechanisms governing
the dynamics of tumor growth.\ One of main reasons for a creation of
non-spatial dynamical models of this nature is related to the fact that they
are described by a system of ordinary differential equations which can be
efficiently investigated by powerful methods of qualitative theory of
ordinary differential equations and dynamical systems theory. In this paper
we examine the dynamics of one cancer growth model proposed in $[5]$ but
possess mutiphase structure, i.e. we consider the dynamical system 
\[
\dot{T}=r_{1}T\left( 1-k_{1}^{-1}T\right) -a_{12}NT-a_{13}TI, 
\]%
\begin{equation}
\dot{N}=r_{2}N\left( 1-k_{2}^{-1}N\right) -a_{21}NT,  \tag{1.1}
\end{equation}%
\[
\dot{I}=\frac{r_{3}}{k_{3}+T}-a_{31}IT-d_{3}I, 
\]%
with multipoint initial condition 
\begin{equation}
T\left( t_{0}\right) =T_{0}+\dsum\limits_{k=1}^{m}\alpha _{1k}T\left(
t_{k}\right) \text{, }N\left( t_{0}\right)
=N_{0}+\dsum\limits_{k=1}^{m}\alpha _{2k}N\left( t_{k}\right) \text{, } 
\tag{1.2}
\end{equation}%
\[
I\left( t_{0}\right) =I_{0}+\dsum\limits_{k=1}^{m}\alpha _{3k}I\left(
t_{k}\right) \text{, }t_{0}\in \left[ 0,\right. \left. T\right) \text{, }%
t_{k}\in \left( 0,T\right) , 
\]%
where $T=T\left( t\right) ,$ $N=N\left( t\right) $, $I=I\left( t\right) $
denote the density of tumor cells, healthy host cells and the effector
immune cells, respectively at the moment $t,$ $\alpha _{jk}$ are real
numbers and $m$ is a natural number. The first term of the first equation
corresponds to the logistic growth of tumor cells in the absence of any
effect from other cells populations with the growth rate of $r_{1}$ and
maximum carrying capacity $k_{1}$. The competition between host cells and
tumor cells $T\left( t\right) $ which results in the loss of the tumor cells
population is given by the term $a_{12}NT$. Next, the parameter $a_{13}$
refers to the tumor cell killing rate by the immune cells $I\left( t\right) $%
. In the second equation, the healthy tissue cells also grow logistically
with the growth rate of $r_{2}$ and maximum carrying capacity $k_{2}$. We
assume that the cancer cells proliferate faster than the healthy cells which
gives $r_{1}>r_{2}$. The tumor cells also inactivate the healthy cells at
the rate of $a_{21}$. The third equation of the model describes the change
in the immune cells population with time $t.$ The first term of the third
equation illustrates the stimulation of the immune system by the tumor cells
with tumor specific antigens. The rate of recognition of the tumor cells by
the immune system depends on the antigenicity of the tumor cells. The model
of the recognition process is given by the rational function which depends
on the number of tumor cells with positive constants $r_{3}$ and $k_{3}$.
The immune cells are inactivated by the tumor cells at the rate of $a_{31}$
as well as they die naturally at the rate $d_{3},$ here we suppose that the
constant influx of the activated effector cells into the tumor
microenvironment is zero.

One of main aim is derivation of sufficient conditions under which the
possible biologically feasible dynamics is local and global stable and a
convergence to one of equilibrium points. Since these equilibrium points
have a biological sense we notice that understanding limit properties of
dynamics of cells populations based on solving problems $(1.1)-\left(
1.2\right) $ may be of an essential interest for the prediction of health
conditions of a patient without a treatment. Note that the local and global
stability properties of $\left( 1.1\right) $ with classical initial
condition were studied in $\left[ 8\right] $ and $\left[ 9\right] $,
respectively. We prove that all orbits are bounded and must converge to one
of several possible equilibrium points. Therefore, the long-term behavior of
an orbit is classified according to the basin of attraction in which it
starts.

By scaling $x_{1}=Tk_{1}^{-1}$, $x_{2}=Nk_{2}^{-1}$, $x_{3}=Ik_{3}^{-1}$, $%
\tilde{t}=r_{1}t$ in $\left( 1.1\right) -\left( 1.2\right) $ and omitting
the tilde notation we obtain the multipoint initial value problem (IVP) 
\[
\dot{x}_{1}=x_{1}\left( 1-x_{1}\right) -a_{12}x_{1}x_{2}-a_{13}x_{1}x_{3}, 
\]%
\begin{equation}
\dot{x}_{2}=r_{2}x_{2}\left( 1-x_{2}\right) -a_{21}x_{1}x_{2},  \tag{1.3}
\end{equation}%
\[
\dot{x}_{3}=\frac{r_{3}x_{1}x_{3}}{x_{1}+k_{3}}-a_{31}x_{1}x_{3}-d_{3}x_{3},%
\text{ }t\in \left[ 0,\right. \left. T\right) , 
\]

\begin{equation}
x_{1}\left( t_{0}\right) =x_{10}+\dsum\limits_{k=1}^{m}\alpha
_{1k}x_{1}\left( t_{k}\right) \text{, }x_{2}\left( t_{0}\right)
=x_{20}+\dsum\limits_{k=1}^{m}\alpha _{2k}x_{2}\left( t_{k}\right) \text{, }
\tag{1.4}
\end{equation}%
\[
x_{3}\left( t_{0}\right) =x_{30}+\dsum\limits_{k=1}^{m}\alpha
_{3k}x_{3}\left( t_{k}\right) \text{, }t_{0}\in \left[ 0,\right. \left.
T\right) \text{, }t_{k}\in \left( 0,T\right) , 
\]%
where $\alpha _{jk}$ are real numbers and $m$ is a natural number such that 
\begin{equation}
x_{j0}+\dsum\limits_{k=1}^{m}\alpha _{jk}x_{j}\left( t_{k}\right) \geq 0%
\text{, }j=1,2,3.  \tag{1.5}
\end{equation}%
Note that, for $\alpha _{j1}=\alpha _{j2}=...\alpha _{jm}=0$ the problem $%
\left( 1.3\right) -\left( 1.4\right) $ turns to be the classical IVP%
\[
\dot{x}_{1}=x_{1}\left( 1-x_{1}\right) -a_{12}x_{1}x_{2}-a_{13}x_{1}x_{3}, 
\]%
\begin{equation}
\dot{x}_{2}=r_{2}x_{2}\left( 1-x_{2}\right) -a_{21}x_{1}x_{2},  \tag{1.6}
\end{equation}%
\[
\dot{x}_{3}=\frac{r_{3}x_{1}x_{3}}{x_{1}+k_{3}}-a_{31}x_{1}x_{3}-d_{3}x_{3},%
\text{ }t\in \left[ 0,T\right] , 
\]%
\[
x_{1}\left( t_{0}\right) =x_{10}\text{, }x_{2}\left( t_{0}\right) =x_{20},%
\text{ }x_{3}\left( t_{0}\right) =x_{30},\text{ }t_{0}\in \left[ 0,\right.
\left. T\right) . 
\]

\begin{center}
\textbf{2}.\textbf{\ Notations and background.}
\end{center}

Consider the multipoint IVP for nonlinear equation%
\begin{equation}
\frac{du}{dt}=f\left( u\right) ,\text{ }t\in \left[ 0,T\right] ,  \tag{2.1}
\end{equation}%
\[
u\left( t_{0}\right) =u_{0}+\dsum\limits_{k=1}^{m}\alpha _{k}u\left(
t_{k}\right) \text{, }t_{0}\in \left[ 0,\right. \left. T\right) \text{, }%
t_{k}\in \left( 0,T\right) , 
\]%
in a Banach space $E$, where $\alpha _{k}$ are complex numbers and $m$ is a
natural number and $u=u\left( t\right) $ is a $E$ valued function. Note
that, for $\alpha _{1}=\alpha _{2}=...\alpha _{m}=0$ the problem $\left(
2.1\right) $ become to be the following local Cauchy problem%
\begin{equation}
\frac{du}{dt}=f\left( u\right) ,\text{ }u\left( t_{0}\right) =u_{0},\text{ }%
t\in \left[ 0,T\right] ,\text{ }t_{0}\in \left[ 0,\right. \left. T\right) 
\text{.}  \tag{2.2}
\end{equation}

We can generalized classical Picard existence theorem for nonlocal nonlinear
problem $\left( 2.1\right) $, i.e. by reasoning as a classical case we obtain

\textbf{Theorem 2.1. }Let $X$ be a Banach space. Suppose that $%
f:X\rightarrow X$ satisfies local Lipschitz condition on a closed ball $\bar{%
B}_{r}(\upsilon _{0})\subset $ $X$, where $r>0$, i.e.%
\[
\left\Vert f\left( u\right) -f\left( \upsilon \right) \right\Vert _{E}\leq
L\left\Vert u-\upsilon \right\Vert _{E} 
\]%
for each $u$, $\upsilon \in \bar{B}_{r}(\upsilon _{0})$, where 
\[
\upsilon _{0}=u_{0}+\dsum\limits_{k=1}^{m}\alpha _{k}u\left( t_{k}\right) 
\]%
and there exists $\delta >0$ such that 
\[
t_{k}\in O_{\delta }\left( t_{0}\right) =\left\{ t\in \mathbb{R}:\left\vert
t-t_{0}\right\vert <\delta \right\} . 
\]

Moreover, let 
\[
M=\sup\limits_{u\in \bar{B}_{r}(\upsilon _{0})}\left\Vert f\left( u\right)
\right\Vert _{X}<\infty . 
\]

Then\ problem $\left( 2.1\right) $ has a unique continuously differentiable
local solution $u(t)$, for $t\in O_{\delta }\left( t_{0}\right) $, where $%
\delta \leq \frac{r}{M}.$

\textbf{Proof.} We rewrite the initial value problem $\left( 2.1\right) $ as
the integral equation 
\[
u=\upsilon _{0}+\dint\limits_{t_{0}}^{t}f\left( u\left( s\right) \right)
ds.\ 
\]

For $0<\eta <\frac{r}{M}$ we define the space 
\[
Y=C\left( \left[ -\eta ,\eta \right] ;\bar{B}_{r}(\upsilon _{0})\right) . 
\]

Let%
\[
Qu=\upsilon _{0}+\dint\limits_{t_{0}}^{t}f\left( u\left( s\right) \right)
ds. 
\]
First, note that if $u\in Y$ then%
\[
\left\Vert Qu-\upsilon _{0}\right\Vert _{X}\leq \left\Vert
\dint\limits_{t_{0}}^{t}f\left( u\left( s\right) \right) ds\right\Vert
_{X}\leq M\eta <r. 
\]

Hence, $Qu\in Y$ so that $Q:Y\rightarrow Y.$ Moreover, for all $u$, $%
\upsilon \in Y$\ we have 
\[
\left\Vert Qu-Q\upsilon \right\Vert _{X}\leq \left\Vert
\dint\limits_{t_{0}}^{t}\left[ f\left( u\left( s\right) \right) -f\left(
\upsilon \left( s\right) \right) \right] ds\right\Vert _{X}\leq 
\]%
\begin{equation}
L_{f}\eta \left\Vert u-\upsilon \right\Vert _{X},  \tag{2.3}
\end{equation}%
where $L_{f}$ is a Lipschitz constant for $f$ on $\bar{B}_{r}(\upsilon _{0})$%
. Hence, if we choose 
\begin{equation}
\eta <\min \left\{ \frac{r}{M},\frac{1}{L_{f}}\right\}  \tag{2.4}
\end{equation}
then $Q$ is a contraction on $Y$ and it has a unique fixed point. Since $%
\eta $ depends only on the Lipschitz constant of $f$ and on the distance $r$
of the initial data from the boundary of $\bar{B}_{r}(\upsilon _{0})$,
repeated application of this result gives a unique local solution defined
for $\left\vert t-t_{0}\right\vert <\frac{r}{M}.$

\textbf{Theorem 2.2. }Let $X$ be a Banach space. Suppose that $%
f:X\rightarrow X$ satisfies global Lipschitz condition, i.e.%
\[
\left\Vert f\left( u\right) -f\left( \upsilon \right) \right\Vert _{E}\leq
L\left\Vert u-\upsilon \right\Vert _{E} 
\]%
for each $u$, $\upsilon \in E$. Moreover, let 
\[
M=\sup\limits_{u\in E}\left\Vert f\left( u\right) \right\Vert _{X}<\infty . 
\]

Then\ problem $\left( 2.1\right) $ has a unique continuously differentiable
local solution $u(t)$, for $\left\vert t-t_{0}\right\vert <\delta $, where $%
\delta \leq \frac{r}{M}.$

\textbf{Proof. }The key point of proof is to show that the constant $\delta $
of Theorem 2.1 can be made independent of the $\upsilon _{0}.$ It is not
hard to see that the independence of $\upsilon _{0}$ comes through the
constant $M$ in therm $\frac{r}{M}$ in $\left( 2.4\right) $. Since in the
current case the Lipschitz condition holds globally, one can choose $r$
arbitrary large. Therefore, for any finite $M$, we can choose $r$ large
enough and by using $\left( 2.3\right) ,$ $\left( 2.4\right) $ we obtain the
assertion.

\begin{center}
\bigskip \textbf{3. Boundedness, invariance of non-negativity, and
dissipativity}
\end{center}

\bigskip In this section, we shall show that the model equations are bounded
with negative divergence, positively (non-negatively) invariant with respect
to a region in $R_{+}^{3}$ and dissipative. As we are interested in
biologically relevant solutions of the system, the next two results show
that the positive octant is invariant and that all trajectories in this
octant are recurrent. Let 
\begin{equation}
B=\left\{ x=\left( x_{1},x_{2},x_{3}\right) \in R_{+}^{3}\text{: }0\leq
x_{i}\leq K_{i}\text{, }i=1,\text{ }2,\text{ }3\right\} .  \tag{3.0}
\end{equation}

\textbf{Theorem 3.1. }Assume\textbf{\ }%
\begin{equation}
d_{3}>1+r_{2},\text{ }r_{i}>0,\text{ }k_{i}>0,\text{ }a_{ij}>0,\text{ }%
r_{3}<k_{3}a_{31}.  \tag{3.1}
\end{equation}%
Then:

(1) $B$ is positively invariant with respect $(1.2)-\left( 1.3\right) ;$

(2) all solutions of problem $(1.2)-\left( 1.3\right) $ with initial values $%
x_{i0}>0$ are uniformly bounded and are attracted into the region $B$;

(3) the system $(1.2)$ is with negative divergence;

(4) the system $(1.2)$ is dissipative.

\textbf{Proof. (}1\textbf{); }Consider the first equation of the system $%
\left( 1.2\right) $:\textbf{\ \ }%
\[
\dot{x}_{1}=x_{1}\left( 1-x_{1}\right) -a_{12}x_{1}x_{2}-a_{13}x_{1}x_{3}. 
\]

By condition $\left( 3.1\right) $ we get 
\[
\dot{x}_{1}<x_{1}\left( 1-x_{1}\right) . 
\]

It is clear that%
\[
x_{1}\left( 1-x_{1}\right) =0,\frac{d}{dx_{1}}\left[ x_{1}\left(
1-x_{1}\right) \right] =1-2x_{1}<0 
\]%
for $x_{1}=1.$\ Thus 
\[
x_{1}\left( t\right) \leq \max \left\{ 1,\text{ }x_{10}-\dsum%
\limits_{k=1}^{m}\alpha _{1k}x_{1}\left( t_{k}\right) \right\} , 
\]%
\[
\dot{x}_{1}<0\text{ for }x_{1}>1. 
\]

Hence, 
\begin{equation}
\limsup\limits_{t\rightarrow \infty }x_{1}\left( t\right) \leq K_{1}=1. 
\tag{3.2 }
\end{equation}

For 
\[
\dot{x}_{2}=r_{2}x_{2}\left( 1-x_{2}\right) -a_{21}x_{1}x_{2}, 
\]%
a similar analysis gives 
\[
x_{2}\left( t\right) \leq \max \left\{ K_{2},\text{ }x_{20}-\dsum%
\limits_{k=1}^{m}\alpha _{2k}x_{2}\left( t_{k}\right) \right\} , 
\]%
\begin{equation}
\limsup\limits_{t\rightarrow \infty }x_{2}\left( t\right) \leq K_{2}. 
\tag{3.3 }
\end{equation}

\bigskip Now consider 
\[
\dot{x}_{3}=\frac{r_{3}x_{1}x_{3}}{x_{1}+k_{3}}-a_{31}x_{1}x_{3}-d_{3}x_{3}. 
\]

From $\left( 3.1\right) $ we have

\[
\dot{x}_{3}<\frac{r_{3}x_{1}x_{3}}{x_{1}+k_{3}}-a_{31}x_{1}x_{3}=x_{1}x_{3}%
\left( \frac{r_{3}}{x_{1}+k_{3}}-a_{31}\right) <0. 
\]

Then by reasoning as the case of $x_{1}$ we deduced 
\[
x_{3}\left( t\right) \leq \max \left\{ K_{3},\text{ }x_{30}-\dsum%
\limits_{k=1}^{m}\alpha _{1k}x_{3}\left( t_{k}\right) \right\} , 
\]%
\begin{equation}
\limsup\limits_{t\rightarrow \infty }x_{3}\left( t\right) \leq K_{3}. 
\tag{3.4}
\end{equation}

Hence, from $\left( 3.2\right) -\left( 3.4\right) $ we obtain (1) and (2)
assumptions.

Now, let us show (3)-(4). Since 
\[
\frac{\partial f_{1}}{\partial x_{1}}+\frac{\partial f_{2}}{\partial x_{2}}+%
\frac{\partial f_{3}}{\partial x_{3}}%
=1-2x_{1}-a_{12}x_{2}+r_{2}-2r_{1}x_{2}-a_{21}x_{1}+ 
\]%
\begin{equation}
\frac{r_{3}x_{1}}{x_{1}+k_{3}}-a_{31}x_{1}-d_{3}=\left( 1+r_{2}-d_{3}\right)
-\left( 2+a_{21}\right) x_{1}-  \tag{3.5}
\end{equation}%
\[
\left( 2r_{1}+a_{12}\right) x_{2}+\left[ \frac{r_{3}}{x_{1}+k_{3}}-a_{31}%
\right] x_{1}. 
\]

By condition $\left( 3.1\right) $ from $\left( 3.5\right) $ we obtain 
\[
\frac{\partial f_{1}}{\partial x_{1}}+\frac{\partial f_{2}}{\partial x_{2}}+%
\frac{\partial f_{3}}{\partial x_{3}}<0\text{ for }x\in B, 
\]%
i.e. the system $(1.2)$ is with negative divergence and is dissipative.

\begin{center}
\textbf{4.1 The equilibria, existence and local stability}
\end{center}

\bigskip The equilibria of system $(1.2)$ are obtained by solving the system
of isocline equations%
\[
x_{1}\left( 1-x_{1}\right) -a_{12}x_{1}x_{2}-a_{13}x_{1}x_{3}=0, 
\]%
\begin{equation}
r_{2}x_{2}\left( 1-x_{2}\right) -a_{21}x_{1}x_{2}=0,  \tag{4.1}
\end{equation}%
\[
\frac{r_{3}x_{1}x_{3}}{x_{1}+k_{3}}-a_{31}x_{1}x_{3}-d_{3}x_{3}=0. 
\]

\bigskip Since we are interested in biologically relevant solutions of $%
\left( 4.1\right) $ we find sufficient conditions under which this system
have positive solutions.

\textbf{Condition 4.1. }Assume:

(1) $r_{3}>d_{3}+a_{31}k_{3};$

(2) $\left( r_{3}-a_{31}k_{3}-d_{3}\right) ^{2}\geq 4d_{3}k_{3}a_{31}.$

\textbf{Lemma 4.1. }Let the Conditi\i n 4.1 hold. Then the system $\left(
1.2\right) $ have the following equilibria points 
\[
E_{0}\left( 0,0,0\right) \text{, }E_{1}\left( 1,0,0\right) \text{, }%
E_{2}\left( 0,1,0\right) \text{, }E\left( \frac{r_{2}}{a_{21}},0,\frac{%
a_{21}-r_{2}}{a_{13}a_{21}}\right) ,\text{ } 
\]%
\begin{equation}
\text{ }E_{ij}\left( x_{1i},x_{2j},x_{3ij}\right) \text{, }i=1,\text{ }2,%
\text{ }j=0\text{, }1\text{, }2,  \tag{4.2}
\end{equation}%
\ where the points $E_{ij}\left( x_{1i},x_{2,j},x_{3,ij}\right) $ will
defined in bellow.

\textbf{Proof. }The possible equilibria are of the form 
\[
E_{0}\left( 0,0,0\right) \text{, }E_{1}\left( 1,0,0\right) \text{, }%
E_{2}\left( 0,1,0\right) ,\text{ }E_{3}\left( \frac{r_{2}}{a_{21}},0,\frac{%
a_{21}-r_{2}}{a_{13}a_{21}}\right) ,\text{ }E_{ij}\left( \bar{x}_{1},\bar{x}%
_{2},\bar{x}_{3}\right) . 
\]

It is clear to see that the points $E_{0}$, $E_{1}$ and $E_{2}$ are
equilibria points. It remain to find the points%
\[
E_{ij}=E_{ij}\left( \bar{x}_{1},\bar{x}_{2},\bar{x}_{3}\right) ,\text{ }%
E\left( \frac{r_{2}}{a_{21}},0,\frac{a_{21}-r_{2}}{a_{13}a_{21}}\right) . 
\]

From the third equation of $\left( 4.1\right) $ for $x_{3}\neq 0$\ we have 
\[
\frac{r_{3}x_{1}}{x_{1}+k_{3}}-a_{31}x_{1}-d_{3}=0, 
\]%
i.e. we have the following square algebraic equation 
\begin{equation}
a_{31}x_{1}^{2}+\left( a_{31}k_{3}+d_{3}-r_{3}\right) x_{1}+d_{3}k_{3}=0. 
\tag{4.3}
\end{equation}%
By solving the equation $\left( 4.3\right) $ \ we get 
\[
x_{1}=\frac{-\left( a_{31}k_{3}+d_{3}-r_{3}\right) \pm \sqrt{D}}{2a_{31}}, 
\]%
where 
\[
D=\left( a_{31}k_{3}+d_{3}-r_{3}\right) ^{2}-4d_{3}k_{3}a_{31}\geq 0. 
\]%
Hence, 
\begin{equation}
x_{11}=\frac{-\left( a_{31}k_{3}+d_{3}-r_{3}\right) +\sqrt{D}}{2a_{31}},%
\text{ }x_{12}=\frac{-\left( a_{31}k_{3}+d_{3}-r_{3}\right) -\sqrt{D}}{%
2a_{31}}.  \tag{4.4}
\end{equation}

From first and second equation of $\left( 4.1\right) $ we have 
\[
\text{(1) }x_{1}=\frac{r_{2}}{a_{21}},\text{ }x_{2}=0,\text{ }x_{3}=\frac{%
a_{21}-r_{2}}{a_{13}a_{21}}, 
\]%
i.e. the point $E\left( \frac{r_{2}}{a_{21}},0,\frac{a_{21}-r_{2}}{%
a_{13}a_{21}}\right) $ is a equilibria point; By taking $\left( 4.4\right) $
in the second equation of $\left( 4.1\right) $ we get 
\[
\text{ (2) }x_{1}>0,\text{ }x_{2}=\frac{r_{2}-a_{21}x_{1}}{r_{2}},\text{ }%
x_{2}\neq 0;\text{ } 
\]

For the case of (2) we get 
\begin{equation}
\text{ }x_{21}=\frac{r_{2}-a_{21}x_{11}}{r_{2}}\text{, }x_{22}=\frac{%
r_{2}-a_{21}x_{12}}{r_{2}}.  \tag{4.5}
\end{equation}

Moreover, by taking $\left( 4.5\right) $ in the first equation of $\left(
4.1\right) $ we obtain%
\begin{equation}
x_{3ij}=\frac{1-x_{1i}-a_{12}x_{2j}}{a_{13}},\text{ }i\text{, }j=1,2. 
\tag{4.6}
\end{equation}

Thus we obtain that the points $\left( 4.2\right) $ are equilibria points
the Jacobian matrix due to liberalization of the system $\left( 1.2\right) $%
, where $x_{1i},$ $x_{2j},$ $x_{3ij}$, $i$, $j=1,2$ are defined by $\left(
4.4\right) -\left( 4.6\right) .$

\textbf{Remark 4.1. }For $a_{21}>r_{2}$ the system $\left( 1.2\right) $ have
the biologically feasible equilibria points 
\[
E_{0}\left( 0,0,0\right) \text{, }E_{1}\left( 1,0,0\right) \text{, }%
E_{2}\left( 0,1,0\right) \text{, }E_{3}\left( \frac{r_{2}}{a_{21}},0,\frac{%
a_{21}-r_{2}}{a_{13}a_{21}}\right) .\text{ } 
\]%
Indeed, our system $\left( 1.2\right) $ describe the biological possess we
have to consider this system in positive domain. So, all roots of $\left(
4.1\right) $ must be positive. For case of (2) $x_{11}>0$ and $x_{12}<0$
when $r_{3}>d_{3}+a_{31}k_{3};$for $r_{3}<d_{3}+a_{31}k_{3}$ both roots $%
x_{11},$ $x_{12}$ are negative; the root $x_{3}$ is positive if $x_{1}<1$,
i.e. \ 
\begin{equation}
r_{3}-\left( a_{31}k_{3}+d_{3}\right) +\sqrt{D}<2a_{31}.  \tag{4.7}
\end{equation}%
The roots $x_{21}$, $x_{22}$ are positive when%
\begin{equation}
r_{3}\geq d_{3}+a_{31}k_{3}+D.  \tag{4.8}
\end{equation}%
Moreover, $x_{3ij},$ $i$, $j=1,2$ positive when 
\begin{equation}
a_{12}<1\text{, }a_{12}a_{21}<r_{2}\text{ or }a_{12}>1,\text{ }%
a_{12}a_{21}>r_{2}.  \tag{4.9}
\end{equation}

Hence, if Condition 4.1 and $\left( 4.7\right) $-$\left( 4.9\right) $ are
satisfied, then the equilibria points $\left( 4.2\right) $ belong to
positive domain%
\[
R_{+}^{3}=\left\{ x\in R^{3}\text{: }x_{i}>0\text{, }i=1,2,3\right\} . 
\]

\textbf{Remark 4.2.} There exist 3 type dead case: (1) for equilibrium $%
E_{0}\left( 0,0,0\right) $ three type cell population are zero; (2) for
point $E_{1}\left( 1,0,0\right) $ tumor cells survive but normal and immune
cells population are zero; (3) for point $E\left( a,0,b\right) $ normal
cells are zero but tumor and immune cells population are survived; (4) $%
E_{2}\left( 0,1,0\right) $-tumor-free and immune free case; in this
category, normal cells survive but tumor and immune cells population are
zero; (5) The equilibrium points $E_{ij}\left( x_{1i},x_{2j},x_{3ij}\right) $%
, $i=1,$ $2,$ $j=0$, $1$, $2$ correspond the cases when tumor, normal and
immune population are survived. We now discuss the (local) linearized
stability of system $\left( 1.2\right) -\left( 1.3\right) $ restricted to a
neighborhood of the equilibrium points $\left( 4.2\right) .$

The Jacobian matrix due to the liberalization of $\left( 1.2\right) $ about
an arbitrary equilibrium point $E\left( x_{1},x_{2},x_{3}\right) $ is given
by%
\begin{equation}
A_{E\left( x_{1},x_{2},x_{3}\right) }=  \tag{4.10}
\end{equation}%
\[
\left[ 
\begin{array}{ccc}
1-2x_{1}-a_{12}x_{2}-a_{13}x_{3} & -a_{12}x_{1} & -a_{13}x_{1} \\ 
-a_{21}x_{2} & r_{2}-2r_{2}x_{2}-a_{21}x_{1} & 0 \\ 
\frac{k_{1}r_{3}x_{3}}{\left( x_{1}+k_{1}\right) ^{2}}-a_{31}x_{3} & 0 & 
\frac{r_{3}x_{1}}{x_{1}+k_{3}}-a_{31}x_{1}-d_{3}%
\end{array}%
\right] . 
\]

The linearized matrices for equilibria points $E_{0}\left( 0,0,0\right) $, $%
E_{1}\left( 1,0,0\right) ,$ $E_{2}\left( 0,1,0\right) $ will be
correspondingly as:%
\[
A_{0}=\left[ 
\begin{array}{ccc}
1 & 0 & 0 \\ 
0 & r_{2} & 0 \\ 
0 & 0 & -d_{3}%
\end{array}%
\right] ,\text{ }A_{1}=\left[ 
\begin{array}{ccc}
-1 & -a_{12} & 0 \\ 
0 & r_{2}-a_{21} & 0 \\ 
0 & 0 & \frac{r_{3}}{1+k_{3}}-a_{31}-d_{3}%
\end{array}%
\right] , 
\]%
\[
A_{2}=\left[ 
\begin{array}{ccc}
1-a_{12} & 0 & 0 \\ 
-a_{21} & -r_{2} & 0 \\ 
0 & 0 & -d_{3}%
\end{array}%
\right] . 
\]

Then we have the linearized matrices for equilibria points $E_{3}\left(
a,0,b\right) $: 
\begin{equation}
A_{3}=A_{E\left( a,0,b\right) }=  \tag{4.11}
\end{equation}%
\[
\left[ 
\begin{array}{ccc}
-a & -a_{12}a & -a_{13}a \\ 
0 & r_{2}-a_{21}a & 0 \\ 
\left( \frac{k_{1}r_{3}}{\left( a+k_{1}\right) ^{2}}-a_{31}\right) b & 0 & 
\frac{r_{3}a}{a+k_{3}}-a_{31}a-d_{3}%
\end{array}%
\right] , 
\]%
where 
\[
a=\frac{r_{2}}{a_{21}},\text{ }b=\frac{a_{21}-r_{2}}{a_{13}a_{21}}. 
\]

\bigskip In a similar way, we find that the linearized matrices $A_{ij}$ for
equilibria points $E_{ij}$ is the following%
\begin{equation}
A_{ij}=A_{E\left( x_{1i},x_{2j},x_{3ij}\right) }=  \tag{4.12}
\end{equation}%
\[
\left[ 
\begin{array}{ccc}
1-2x_{1i}-a_{12}x_{2j}-a_{13}x_{3ij} & -a_{12}x_{1i} & -a_{13}x_{1i} \\ 
-a_{21}x_{2} & r_{2}-2r_{2}x_{2j}-a_{21}x_{1i} & 0 \\ 
\frac{k_{1}r_{3}x_{3ij}}{\left( x_{1i}+k_{1}\right) ^{2}}-a_{31}x_{3ij} & 0
& \frac{r_{3}x_{1i}}{x_{1i}+k_{3}}-a_{31}x_{1i}-d_{3}%
\end{array}%
\right] . 
\]

\begin{center}
\textbf{5. local stability analysis of points }$E_{0}\left( 0,0,0\right) $
and $E_{1}\left( 0,1,0\right) $
\end{center}

In this section we show the following \ result:

\textbf{Theorem 5.1. (}1) The point $E_{0}$ is unstable point for the
linearized system of $\left( 1.2\right) ;$ (2) The point $E_{1}$ is locally
asymptotically stable point for the linearized system of $\left( 1.2\right) $
when $a_{21}>r_{2}$, $a_{31}+d_{3}>\frac{r_{3}}{1+k_{3}}$ and is an unstable
point when $a_{21}<r_{2}$ and $a_{31}+d_{3}<\frac{r_{3}}{1+k_{3}}$;

(3) The point $E_{2}$ is locally asymptotically stable point for linearized
system of $\left( 1.2\right) $ when $a_{12}>1$ and is an unstable point when 
$a_{12}<1.$

\textbf{Proof. }Indeed, the eigenvalues of $A_{E_{0}\left( 0,0,0\right) }$
are $1,$ $r_{2},$ $-d_{3},$ eigenvalues of $A_{1}\left( 1,0,0\right) $ are 
\[
-1,\text{ }r_{2}-a_{21},\text{ }\frac{r_{3}}{1+k_{3}}-a_{31}-d_{3}; 
\]
and eigenvalues of $E_{2}\left( 0,1,0\right) $ are 
\[
1-a_{12},\text{ }-r_{2},\text{ }-d_{3}. 
\]

Since $1$, $r_{2}>0$, then by $\left[ \text{5, Theorem 8.12}\right] $ $E_{0}$
is unstable point. Due to negativity of $-1,$ $r_{2}-a_{21},$ $\frac{r_{3}}{%
1+k_{3}}-a_{31}-d_{3}$\ and $1-a_{12}$, $-r_{2}$, $-d_{3}$ by $\left[ \text{%
5, Theorem 8.12}\right] $\ we get that $E_{1}$ and $E_{2}$ are locally
asymptotically stable points.

Now, consider the Jacobian matrices $A_{ij}=\left[ b_{km}\right] ,$ $k$, $%
m=1,2,3$ to the linearized system of $\left( 1.2\right) $ on points $E_{ij}$
defined by $\left( 4.10\right) .$ Let 
\[
b_{11}=b_{11}\left( ij\right) =1-2x_{1i}-a_{12}x_{2j}-a_{13}x_{3ij},\text{ }%
b_{21}= 
\]

\[
b_{21}\left( ij\right) =-a_{21}x_{2},\text{ }b_{22}=b_{22}\left( ij\right)
=r_{2}-2r_{2}x_{2j}-a_{21}x_{1i}, 
\]

\[
\text{ }b_{13}=b_{13}\left( ij\right) =-a_{13}x_{1i}\text{, }%
b_{31}=b_{31}\left( ij\right) = 
\]%
\[
\frac{k_{1}r_{3}x_{3ij}}{\left( x_{1i}+k_{1}\right) ^{2}}-a_{31}x_{3ij}\text{%
, }b_{33}=b_{33}\left( ij\right) =\frac{r_{3}x_{1i}}{x_{1i}+k_{3}}%
-a_{31}x_{1i}-d_{3}. 
\]

Here we prove the following results:

\textbf{Theorem 5.2. }Assume\textbf{\ }the following conditions are
satisfied:

(1) $\left( a_{31}k_{3}+d_{3}-r_{3}\right) ^{2}\geq 4d_{3}a_{31}$;

(2) $\frac{r_{3}x_{1i}}{x_{1i}+k_{3}}<a_{31}x_{1i}+d_{3}$ and $%
2x_{1i}+a_{12}x_{2j}+a_{13}x_{3ij}+a_{31}x_{1i}>1$;

(3) $b_{11}b_{33}-b_{13}b_{31}>0.$

Then the points $E_{ij}\left( x_{1i},x_{2j},x_{3ij}\right) $ are locally
asymptotically stable points to the linearized system of $\left( 1.2\right) $%
.

\bigskip \textbf{Proof. }Eigne value of $A_{ij}$ are find as roots of the
equations%
\[
\left\vert A_{ij}-\lambda \right\vert =\left\vert 
\begin{array}{ccc}
b_{11}\left( ij\right) -\lambda & b_{12}\left( ij\right) & b_{13}\left(
ij\right) \\ 
b_{21}\left( ij\right) & b_{22}\left( ij\right) -\lambda & 0 \\ 
b_{31}\left( ij\right) & 0 & b_{33}\left( ij\right) -\lambda%
\end{array}%
\right\vert =0, 
\]%
i.e.,%
\[
\left\vert A_{ij}-\lambda \right\vert =\left( b_{11}-\lambda \right) \left(
b_{22}-\lambda \right) \left( b_{33}-\lambda \right) -b_{13}b_{31}\left(
b_{22}-\lambda \right) - 
\]%
\begin{equation}
b_{21}b_{13}\left( b_{33}-\lambda \right) =0.  \tag{5.1}
\end{equation}

From $\left( 4.10\right) $ we get that if $\lambda =\lambda _{1}=b_{33}$,
then the equation $\left( 5.1\right) $ reduced to 
\[
\left( b_{22}-\lambda \right) \left[ \left( b_{11}-\lambda \right) \left(
b_{33}-\lambda \right) -b_{13}b_{31}\right] =0, 
\]%
i.e., 
\begin{equation}
\lambda ^{2}-\left( b_{11}+b_{33}\right) \lambda
+b_{11}b_{33}-b_{13}b_{31}=0.  \tag{5.2}
\end{equation}

\bigskip By solving $\left( 5.2\right) $ we obtain 
\begin{equation}
\lambda _{2}=\frac{\left( b_{11}+b_{33}\right) +\sqrt{\left(
b_{11}-b_{33}\right) ^{2}+4b_{13}b_{31}}}{2},  \tag{5.3}
\end{equation}%
\[
\lambda _{3}=\frac{\left( b_{11}+b_{33}\right) -\sqrt{\left(
b_{11}-b_{33}\right) ^{2}+4b_{13}b_{31}}}{2}; 
\]

Since $b_{\mu \nu }=b_{\mu \nu }\left( ij\right) $, then eigne values $%
\left( \lambda _{1},\text{ }\lambda _{2}\text{, }\lambda _{3}\right) $ will
dependent on $i,$ $j$, i.e., the sets $\left( \lambda _{1}\left( ij\right) ,%
\text{ }\lambda _{2}\left( ij\right) \text{, }\lambda _{3}\left( ij\right)
\right) $ are eigne values of matrices $A_{ij}$, respectively. Hence, by
assumption (2) roots $\lambda _{2}$, $\lambda _{3}$ of $\left( 5.2\right) $
are real and 
\[
\lambda _{1}=b_{33}\left( ij\right) =\frac{r_{3}x_{1i}}{x_{1i}+k_{3}}%
-a_{31}x_{1i}-d_{3}<0, 
\]

\begin{equation}
b_{11}+b_{33}=1-2x_{1i}-a_{12}x_{2j}-a_{13}x_{3ij}+\frac{r_{3}x_{1i}}{%
x_{1i}+k_{3}}-a_{31}x_{1i}-d_{3}<0.  \tag{5.4}
\end{equation}

Then from assumption (3) and from $\left( 5.2\right) -\left( 5.4\right) $ we
get that $\lambda _{k}<0,$ $k=1,2,3$ when 
\[
\left( b_{11}-b_{33}\right) ^{2}\geq -4b_{13}b_{31}. 
\]%
But $\lambda _{1}$ is a negative number and $\lambda _{2}$, $\lambda _{3}$
are complex numbers with negative real part, when 
\[
\left( b_{11}-b_{33}\right) ^{2}<-4b_{13}b_{31}. 
\]%
That is, under assumption (1)-(3) the points $E_{ij}$ are locally
asymptotically stable points of $\left( 1.2\right) $.

\textbf{Remark 5.1. }In view of assumptions (2), if \ 
\[
\left[ a_{31}-\frac{k_{1}r_{3}}{\left( x_{1i}+k_{1}\right) ^{2}}\right]
x_{1i}x_{3ij}\geq 0, 
\]%
then$\ $condition (3) holds.

\textbf{Remark 5.2.} It is clear to see that the assumptions 
\[
\left( b_{11}-b_{33}\right) ^{2}\geq -4b_{13}b_{31},\left(
b_{11}-b_{33}\right) ^{2}<-4b_{13}b_{31} 
\]%
are satisfied under condition on coefficients of $\left( 1.2\right) $. One
can fined this condition according its.

\textbf{Theorem 5.3. }Assume\textbf{\ }the following conditions are
satisfied:

(1) $\left( a_{31}k_{3}+d_{3}-r_{3}\right) ^{2}\geq 4d_{3}a_{31}$;

(2) $a_{31}x_{1i}+d_{3}<\frac{r_{3}x_{1i}}{x_{1i}+k_{3}}$ and $%
2x_{1i}+a_{12}x_{2j}+a_{13}x_{3ij}<1$;

(3) $b_{11}b_{33}-b_{13}b_{31}>0.$

Then the points $E_{ij}\left( x_{1i},x_{2j},x_{3ij}\right) $ are locally
unstable points for the linearized system of $\left( 1.2\right) $.

\textbf{Proof. }Indeed, by assumption $(2)$ roots $\lambda _{2}$, $\lambda
_{3}$ of $\left( 4.8\right) $ are real and 
\begin{equation}
\lambda _{1}=b_{33}\left( ij\right) =\frac{r_{3}x_{1i}}{x_{1i}+k_{3}}%
-a_{31}x_{1i}-d_{3}>0  \tag{5.5}
\end{equation}

\[
b_{11}+b_{33}=1-2x_{1i}-a_{12}x_{2j}-a_{13}x_{3ij}+\frac{r_{3}x_{1i}}{%
x_{1i}+k_{3}}-a_{31}x_{1i}-d_{3}>0. 
\]

From assumption (3), from $\left( 5.5\right) $ and $\left( 5.2\right)
-\left( 5.3\right) $ we get that if $\left( b_{11}-b_{33}\right) ^{2}\geq
-4b_{13}b_{31},$ then $\lambda _{k}>0,$ $k=1$, $2$, $3.$ But if $\left(
b_{11}-b_{33}\right) ^{2}<-4b_{13}b_{31}$, then $\lambda _{1}$ is a positive
number and $\lambda _{2}$, $\lambda _{3}$ are complex numbers with positive
real part, i.e, under this assumptions the points $E_{ij}\left(
x_{1i},x_{2j},x_{3ij}\right) $ are locally unstable points of the system $%
\left( 1.2\right) .$

\textbf{Theorem 5.4. }Assume\textbf{\ }the following conditions are
satisfied:

(1) $\left( a_{31}k_{3}+d_{3}-r_{3}\right) ^{2}\geq 4d_{3}a_{31}$;

(2) $b_{11}b_{33}-b_{13}b_{31}<0.$

Then the points $E_{ij}\left( x_{1i},x_{2j},x_{3ij}\right) $ are saddle
points for the linearized system of $\left( 1.2\right) $.

\textbf{Proof. }Indeed, by assumption $(2)$ eigenvalues $\lambda _{2}$, $%
\lambda _{3}$ are real and of opposite sign, i.e. $E_{ij}\left(
x_{1i},x_{2j},x_{3ij}\right) $ are saddle points of the system $\left(
1.2\right) .$

\textbf{Theorem 5.5. }Let $\left( 4.2\right) $ holds $b_{11}=b_{33}=0$ and $%
b_{13}b_{22}b_{31}<0.$ Then $TrA_{ij}=\lambda _{1}+\lambda _{2}+\lambda
_{3}=0$ and Det $A_{ij}>0$, i.e. the points $E\left(
x_{1i},x_{2j},x_{3ij}\right) $ are centers to linearized system of $\left(
1.2\right) .$

\textbf{Proof. }For $b_{11}+b_{33}=0$ from $\left( 5.2\right) -\left(
5.4\right) $ we get that eigenvalues $\lambda _{2}$, $\lambda _{3}$ are $\mp
\psi ,$ $\omega \in \mathbb{R}$ when $\left( b_{11}-b_{33}\right) ^{2}\geq
-4b_{13}b_{31}$ and $\lambda _{2}$, $\lambda _{3}$ are $\mp i\psi $ when $%
\left( b_{11}-b_{33}\right) ^{2}<-4b_{13}b_{31}.$ Hence, for $%
b_{11}=b_{33}=0 $ we obtain 
\[
TrA_{ij}=\lambda _{1}+\lambda _{2}+\lambda _{3}=0. 
\]%
Moreover,%
\[
\text{Det }A_{ij}=-b_{13}b_{22}b_{31}>0, 
\]%
i.e. we obtain the assertion.

\textbf{Remark 5.3. }For $b_{11}b_{33}-b_{13}b_{31}=0$ we obtain 
\[
\lambda _{2}=0,\text{ }\lambda _{3}=\frac{\left( b_{11}+b_{33}\right) }{2}. 
\]

Now, consider the equilibria points $E_{3}\left( a,0,b\right) .$

\textbf{Condition 5.6. }Assume:

(1) $\left( a_{31}k_{3}+d_{3}-r_{3}\right) ^{2}\geq 4d_{3}a_{31}$;

(2) $\frac{a_{21}r_{3}}{r_{2}+a_{21}k_{3}}<a_{31}+\frac{a_{21}d_{3}}{r_{2}}%
+1;$

(3) $a_{31}+1>\frac{r_{3}}{k_{3}},$ $a=\frac{r_{2}}{a_{21}},$ $b=\frac{%
a_{21}-r_{2}}{a_{13}a_{21}},$ $a_{21}>\frac{r_{2}}{a};$

(4) $c_{11}c_{33}-c_{13}c_{31}>0.$

\textbf{Theorem 5.6. }Assume\textbf{\ }the Condition 5.6 hold. Then the
point $E_{3}\left( a,0,b\right) $ is a locally asymptotically stable point
for linearized system of $\left( 1.2\right) $.

\bigskip \textbf{Proof. }Eigne value of matrices $A_{3}$ are find as roots
of the equations%
\[
\left\vert A_{3}-\lambda \right\vert =\left\vert 
\begin{array}{ccc}
c_{11}-\lambda & c_{12} & c_{13} \\ 
0 & c_{22}-\lambda & 0 \\ 
c_{31} & 0 & c_{33}-\lambda%
\end{array}%
\right\vert =0, 
\]%
i.e.,%
\begin{equation}
\left\vert A_{3}-\lambda \right\vert =\left( c_{11}-\lambda \right) \left(
c_{22}-\lambda \right) \left( c_{33}-\lambda \right) -c_{13}c_{31}\left(
c_{22}-\lambda \right) =  \tag{5.6}
\end{equation}%
\[
\left( c_{22}-\lambda \right) \left[ \lambda ^{2}-\left(
c_{11}+c_{33}\right) \lambda +c_{11}c_{33}-c_{13}c_{31}\right] =0, 
\]%
where 
\[
c_{11}=-a\text{, }c_{12}=-a_{12}a,\text{ }c_{13}=-a_{13}a,\text{ }%
c_{22}=r_{2}-a_{21}a\text{,} 
\]%
\[
c_{31}=\left( \frac{k_{1}r_{3}}{\left( a+k_{1}\right) ^{2}}-a_{31}\right) b,%
\text{ }c_{33}=\frac{r_{3}a}{a+k_{3}}-a_{31}a-d_{3}. 
\]

From $\left( 5.6\right) $ we get that $\lambda =\lambda _{1}=c_{22}$ is one
of eigne value of $A_{3}$ and rest eigne values are the root of the equation 
\begin{equation}
\left[ \lambda ^{2}-\left( c_{11}+c_{33}\right) \lambda
+c_{11}c_{33}-c_{13}c_{31}\right] =0.  \tag{5.7}
\end{equation}

\bigskip By solving $\left( 5.7\right) $ we obtain 
\begin{equation}
\lambda _{2}=\frac{\left( c_{11}+c_{33}\right) +\sqrt{\left(
c_{11}-c_{33}\right) ^{2}+4c_{13}c_{31}}}{2},  \tag{5.8}
\end{equation}%
\[
\lambda _{3}=\frac{\left( c_{11}+c_{33}\right) -\sqrt{\left(
c_{11}-c_{33}\right) ^{2}+4c_{13}c_{31}}}{2}; 
\]

The set $\left( \lambda _{1},\text{ }\lambda _{2}\text{, }\lambda
_{3}\right) $ are eigne values of matrices $A_{3}$, respectively. Hence, by
assumption (3) and (2) we get 
\begin{equation}
\lambda _{1}=r_{2}-a_{21}a<0,\text{ }c_{11}+c_{33}=-a+\frac{r_{3}a}{a+k_{3}}%
-a_{31}a-d_{3}<0.  \tag{5.9}
\end{equation}

Then from $\left( 5.6\right) -\left( 5.9\right) $ we get that $\lambda
_{k}<0,$ $k=1,$ $2$, $3$ when%
\begin{equation}
\left( c_{11}-c_{33}\right) ^{2}\geq -4c_{13}c_{31}.  \tag{5.10}
\end{equation}%
But $\lambda _{1}$ is negative and $\lambda _{2}$, $\lambda _{3}$ are
complex numbers with negative real part when 
\begin{equation}
\left( c_{11}-c_{33}\right) ^{2}<-4c_{13}c_{31}.  \tag{5.11}
\end{equation}%
That is, under assumption (1)-(3) the point $E_{3}\left( a,0,b\right) $ is a
locally asymptotically stable point of the system $\left( 1.2\right) $.

\textbf{Theorem 5.7.}{\ }Assume\textbf{\ }the the following conditions are
satisfied:

(1) $\left( a_{31}k_{3}+d_{3}-r_{3}\right) ^{2}\geq 4d_{3}a_{31}$;

(2) $\frac{a_{21}r_{3}}{r_{2}+a_{21}k_{3}}>a_{31}+\frac{a_{21}d_{3}}{r_{2}}%
+1,$ $a_{21}<\frac{r_{2}}{a};$

(3) $c_{11}c_{33}-c_{13}c_{31}>0.$

Then the point $E_{3}\left( a,0,b\right) $ is a locally unstable point for $%
\left( 1.2\right) $

\textbf{Proof. }Indeed, by assumption (2) roots $\lambda _{2}$, $\lambda
_{3} $ of $\left( 4.16\right) $ are real and 
\[
\lambda _{1}=c_{22}=r_{2}-a_{21}a>0, 
\]

\begin{equation}
c_{11}+c_{33}=-a+\frac{r_{3}a}{a+k_{3}}-a_{31}a-d_{3}>0.  \tag{5.12}
\end{equation}

Then from assumption (3) and from $\left( 5.7\right) $, $\left( 5.8\right) $%
, $\left( 4.12\right) $ we get that $\lambda _{k}<0,$ $k=1,$ $2$, $3$ when $%
\left( c_{11}-c_{33}\right) ^{2}>-4c_{13}c_{31}.$ But if $\left(
c_{11}-c_{33}\right) ^{2}<-4c_{13}c_{31}$, then $\lambda _{1}<0$ and $%
\lambda _{2}$, $\lambda _{3}$ are complex numbers with negative real part.
That is, under assumption (1)-(3) the points $E\left( a,0,b\right) $ are
locally unstable points for $\left( 1.2\right) $.

\textbf{Theorem 5.8. }Let the assumption $\left( 4.2\right) $ hold and $%
b_{11}b_{33}-b_{13}b_{31}>0$. Then, the dimensions of the stable manifold $%
W^{+}$ and unstable manifold $W^{-}$ are given, respectively, by 
\[
\text{Dim }W^{+}\left( E_{ij}\left( x_{1i},x_{2j},x_{3ij}\right) \right) =2,%
\text{ Dim }\left( W^{-}E_{ij}\left( x_{1i},x_{2j},x_{3ij}\right) \right)
=2. 
\]

\textbf{Proof. }Let we solve the the following matrix equation%
\[
A_{ij}x=\lambda x, 
\]%
i.e. consider the system of homogenous linear equation 
\[
\left( b_{11}-\lambda \right) x_{1}+b_{12}x_{2}+b_{13}x_{3}=0, 
\]%
\begin{equation}
b_{21}x_{1}+\left( b_{22}-\lambda \right) x_{2}=0,  \tag{5.13}
\end{equation}%
\[
b_{31}x_{1}+\left( b_{33}-\lambda \right) x_{3}=0, 
\]%
where 
\[
b_{\mu \nu }=b_{\mu \nu }\left( ij\right) \text{, }\mu \text{, }\nu =1,2,3%
\text{, }i=1,2,\text{ }j=0,1,2. 
\]%
By solving of $\left( 5.13\right) $ we obtain that the subspaces 
\begin{equation}
B\left( x_{1i},x_{2j},x_{3ij}\right) =\left\{ x=\left( 0,\text{ }-\frac{%
b_{21}}{b_{22}-\lambda },\text{ }-\frac{b_{31}}{b_{33}-\lambda }\right) a\in
R^{3}\text{ }\right\}  \tag{5.14}
\end{equation}%
are eigne subspaces of matrices $A_{ij}$, where $a$ is arbitrary number from 
$\mathbb{R}.$

Let the assumption (2) of Theorem 5.2 are hold, then we get that 
\[
B\left( x_{1i},x_{2j},x_{3ij}\right) =W^{+}\left( E_{ij}\left(
x_{1i},x_{2j},x_{3ij}\right) \right) ; 
\]%
if the assumption (2) of Theorem 5.3 are hold, then we have 
\[
B\left( x_{1i},x_{2j},x_{3ij}\right) =W^{-}\left( E_{ij}\left(
x_{1i},x_{2j},x_{3ij}\right) \right) . 
\]%
In view of $\left( 5.14\right) $ we obtain the assertion.

In a similar way we obtain

\textbf{Theorem 5.9. }Let the assumption $\left( 4.2\right) $ hold and $%
b_{11}b_{33}-b_{13}b_{31}\leq 0.$Then, the dimensions of the hyperbolic
saddle manifold $W^{0}$ are given and 
\[
\text{Dim }W^{0}\left( E_{ij}\left( x_{1i},x_{2j},x_{3ij}\right) \right) =2. 
\]

\bigskip \textbf{Definition 5.1.} A set $A\subset S$ is called a strong
attractor with respect to $S$ if%
\[
\limsup\limits_{t\rightarrow \infty }\rho \left( u\left( t\right) ,A\right)
=0, 
\]%
where $u\left( t\right) $ is an orbit such that $u\left( t_{0}\right)
-\dsum\limits_{k=1}^{m}\alpha _{k}u\left( t_{k}\right) \in $ $S$ and $\rho $
is the Euclidean distance function.

\textbf{Lemma 5.1. }The $B$ is a strong attractor set with respect to $%
R_{+}^{3}.$

\textbf{Proof. }The proof is done using standard comparison as in Theorem
3.1.

\begin{center}
\textbf{6. Global stability of equilibria points}
\end{center}

\bigskip In this section, we derive the sufficient conditions for the global
stability to system $\left( 1.2\right) $ on the domain $B\subset R_{+}^{3}$
defined by $\left( 3.0\right) $. Consider the equilibria point $E\left( \bar{%
x}_{1},\bar{x}_{2},\bar{x}_{3}\right) .$

\bigskip \textbf{Theorem 6.1. }Assume the following conditions are satisfied:

(1) $\bar{x}=\left( \bar{x}_{1},\bar{x}_{2},\bar{x}_{3}\right) \in B$ is a
local asymptotic stable point of linearized system $\left( 1.2\right) ;$

(2) $a_{13}d_{3}>r_{3};$

(3) $\left( a_{31}k_{3}+d_{3}-r_{3}\right) ^{2}\geq 4d_{3}a_{31}$.

Then the equilibria solution $\bar{x}=\left( \bar{x}_{1},\bar{x}_{2},\bar{x}%
_{3}\right) $ is global stable in the sense of Lyapunov.

\textbf{Proof. }Consider the candidate of Lyapunov function $V\left(
x\right) $ defined by 
\begin{equation}
V\left( x\right) =\dsum\limits_{i,j=1}^{3}d_{ij}\left( x_{i}-\bar{x}%
_{i}\right) \left( x_{j}-\bar{x}_{j}\right) \text{, }d_{ij}\in \mathbb{R}%
\text{ and }a_{ij}=a_{ji}.  \tag{6.1}
\end{equation}

It is clear that 
\[
V\left( x\right) =\left( x-\bar{x}\right) ^{T}A\left( x-\bar{x}\right) = 
\]

\[
\left[ x_{1}-\bar{x}_{1},x_{2}-\bar{x}_{2},x_{3}-\bar{x}_{3}\right] \left[ 
\begin{array}{ccc}
d_{11} & d_{12} & d_{13} \\ 
d_{21} & d_{22} & d_{23} \\ 
d_{31} & d_{32} & d_{33}%
\end{array}%
\right] \left[ 
\begin{array}{c}
x_{1}-\bar{x}_{1} \\ 
x_{2}-\bar{x}_{2} \\ 
x_{3}-\bar{x}_{3}%
\end{array}%
\right] = 
\]%
\begin{equation}
\dsum\limits_{k=1}^{3}d_{kk}\left( x_{k}-\bar{x}_{k}\right)
^{2}+\dsum\limits_{i,j=1}^{3}2d_{ij}\left( x_{i}-\bar{x}_{i}\right) \left(
x_{j}-\bar{x}_{j}\right) ,  \tag{6.2}
\end{equation}%
where 
\[
d_{ij}=d_{ji},\text{ }d_{ij}\in \mathbb{R},\text{ }x\in R^{3}. 
\]

It is known that the quadratic forma defined by $\left( 5.1\right) $ is
positive defined, when 
\[
d_{11}>0,\text{ }d_{11}d_{22}-d_{12}^{2}>0,\text{ Det }%
A=d_{11}d_{22}d_{33}+2d_{12}d_{13}d_{23}- 
\]%
\[
\left( d_{13}^{2}d_{22}+d_{11}d_{23}^{2}+d_{12}^{2}d_{33}\right)
=d_{33}\left( d_{11}d_{22}-d_{12}^{2}\right) +2d_{12}d_{13}d_{23}- 
\]%
\[
\left( d_{13}^{2}d_{22}+d_{11}d_{23}^{2}\right) >0. 
\]%
Hence, if we assume $d_{11}>0,$ $d_{33}>0,$ $d_{11}d_{22}-d_{12}^{2}>0$ and $%
d_{13}^{2}d_{22}+d_{11}d_{23}^{2}<2d_{12}d_{13}d_{23}$ then $V\left(
x\right) >0.$

Moreover, the orbital derivative of $V\left( x\right) $ with respect to
system $\left( 1.2\right) $\ is given by

\[
L_{t}V=\dot{V}\left( x\right) =\dsum\limits_{k=1}^{3}\frac{\partial V}{%
\partial x_{k}}\frac{dx_{k}}{dt}=\dsum\limits_{k,j=1}^{3}2d_{kj}\left( x_{j}-%
\bar{x}_{j}\right) \frac{dx_{k}}{dt}= 
\]

\[
-\left[ \left( 2d_{11}+2d_{12}a_{12}\right) x_{1}^{2}+\left(
2d_{12}+2d_{22}r_{2}\right) x_{2}^{2}+2d_{13}a_{13}x_{3}^{2}\right] -
\]%
\[
2d_{11}\left[ a_{12}x_{1}x_{2}+a_{13}x_{1}x_{3}\right] -2d_{12}\left[
x_{1}x_{2}+a_{13}x_{2}x_{3}\right] -2d_{13}\left[ x_{1}x_{3}+a_{12}x_{2}x_{3}%
\right] -
\]%
\begin{equation}
\left[ \left( 2d_{12}r_{2}+2d_{22}a_{21}\right)
x_{1}x_{2}+2d_{33}r_{2}x_{2}x_{3}+2d_{33}a_{21}x_{1}x_{3}\right] +  \tag{6.3}
\end{equation}%
\[
\left[ 2d_{11}\left( 1+\bar{x}_{1}\right) +2d_{12}\bar{x}_{2}+2d_{13}\bar{x}%
_{3}+2d_{12}r_{2}+2a_{21}\left( d_{12}\bar{x}_{1}+d_{22}\bar{x}_{2}+d_{23}%
\bar{x}_{3}\right) \right] x_{1}+
\]%
\[
\left[ 2d_{11}a_{12}\bar{x}_{1}+2d_{12}\bar{x}_{2}+2d_{12}a_{12}\bar{x}%
_{2}+2d_{12}r_{2}\bar{x}_{1}+2d_{22}r_{2}\left( 1+\bar{x}_{2}\right)
+2d_{33}r_{2}\bar{x}_{3}\right] x_{2}+
\]%
\[
\left[ 2d_{11}a_{13}\bar{x}_{1}+2d_{12}a_{13}\bar{x}%
_{2}+2d_{13}+2d_{13}a_{13}\bar{x}_{3}\right] x_{3}-
\]%
\[
2r_{2}\left[ d_{12}\bar{x}_{1}+d_{22}\bar{x}_{2}+d_{33}\bar{x}_{3}\right]
+2d_{12}r_{2}\left( x_{1}+\bar{x}_{1}x_{2}\right) +
\]%
\[
-2\left[ d_{11}\bar{x}_{1}+d_{12}\bar{x}_{2}+d_{13}\bar{x}_{3}+r_{2}\left(
d_{12}\bar{x}_{1}+d_{22}\bar{x}_{2}+d_{32}\bar{x}_{3}\right) \right] +
\]%
\[
\left[ 2\left( d_{31}\left( x_{1}-\bar{x}_{1}\right) +d_{32}\left( x_{2}-%
\bar{x}_{2}\right) +d_{33}\left( x_{3}-\bar{x}_{3}\right) \right) \right] %
\left[ \frac{r_{3}x_{1}}{x_{1}+k_{3}}-a_{13}x_{1}-d_{3}\right] .
\]

\bigskip Since for $x\in \Omega _{K},$ 
\[
-2d_{11}\left[ a_{12}x_{1}x_{2}+a_{13}x_{1}x_{3}\right] \leq 0,\text{ } 
\]%
in view of inequalities 
\begin{equation}
2ab\leq a^{2}+b^{2},\text{ }x_{1}^{2}+x_{2}^{2}\leq \left\Vert x\right\Vert
^{2},\text{ }x_{2}^{2}+x_{3}^{2}\leq \left\Vert x\right\Vert ^{2}  \tag{6.4}
\end{equation}%
$\left( 6.3\right) $ holds if 
\begin{equation}
\frac{r_{3}x_{1}}{x_{1}+k_{3}}-a_{13}x_{1}-d_{3}\leq 0,  \tag{6.5 }
\end{equation}%
\[
-\left[ \left( 2d_{11}+2d_{12}a_{12}\right) x_{1}^{2}+\left(
2d_{12}+2d_{22}r_{2}\right) x_{2}^{2}+2d_{13}a_{13}x_{3}^{2}\right] + 
\]%
\[
\left\vert d_{12}\right\vert \left( x_{1}^{2}+x_{2}^{2}\right) +\left\vert
d_{12}\right\vert a_{13}\left( x_{2}^{2}+x_{3}^{2}\right) +\left\vert
d_{13}\right\vert \left( x_{1}^{2}+x_{3}^{2}\right) +\left\vert
d_{13}\right\vert a_{13}\left( x_{2}^{2}+x_{3}^{2}\right) + 
\]%
\[
\left( \left\vert d_{12}\right\vert r_{2}+\left\vert d_{22}\right\vert
a_{21}\right) \left( x_{1}^{2}+x_{2}^{2}\right) +\left\vert
d_{33}\right\vert r_{2}\left( x_{2}^{2}+x_{3}^{2}\right) +\left\vert
d_{33}\right\vert a_{21}\left( x_{1}^{2}+x_{3}^{2}\right) \leq 
\]%
\begin{equation}
-\left[ \left( 2d_{11}+2d_{12}a_{12}\right) x_{1}^{2}+\left(
2d_{12}+2d_{22}r_{2}\right) x_{2}^{2}+2d_{13}a_{13}x_{3}^{2}\right] +\eta
\left\vert x\right\vert ^{2}<0\text{,}  \tag{6.6}
\end{equation}

\begin{equation}
\eta _{1}x_{1}+\eta _{2}x_{2}+\eta _{1}x_{3}<0,\text{ }d_{11}\bar{x}%
_{1}+d_{12}\bar{x}_{2}+d_{13}\bar{x}_{3}<0,  \tag{6.7}
\end{equation}%
where 
\[
\eta =\max \left\{ \left( \left\vert d_{12}\right\vert r_{2}+\left\vert
d_{22}\right\vert a_{21}\right) \text{, }\left\vert d_{33}\right\vert \text{%
, }d_{33}r_{2}\right\} ,\text{ } 
\]

\[
\eta _{1}=2d_{11}\left( 1+\bar{x}_{1}\right) +2d_{12}\bar{x}_{2}+2d_{13}\bar{%
x}_{3}+2d_{12}r_{2}+ 
\]

\begin{equation}
2a_{21}\left( d_{12}\bar{x}_{1}+d_{22}\bar{x}_{2}+d_{23}\bar{x}_{3}\right) ,
\tag{6.8}
\end{equation}%
\[
\eta _{2}=2d_{11}a_{12}\bar{x}_{1}+2d_{12}\bar{x}_{2}+2d_{12}a_{12}\bar{x}%
_{2}+2d_{12}r_{2}\bar{x}_{1}+ 
\]%
\[
2d_{22}r_{2}\left( 1+\bar{x}_{2}\right) +2d_{33}r_{2}\bar{x}_{3}, 
\]%
\[
\eta _{3}=2d_{11}a_{13}\bar{x}_{1}+2d_{12}a_{13}\bar{x}%
_{2}+2d_{13}+2d_{13}a_{13}\bar{x}_{3}. 
\]

By assumption (2), $\left( 6.5\right) $ holds for all $x\in $ $\Omega _{K}.$
The inequality $\left( 6.5\right) $ satisfied for all $x\in $ $\Omega _{K},$
when%
\begin{equation}
\eta <\min \left\{ \left( 2d_{11}+2\left\vert d_{12}\right\vert
a_{12}\right) ,\text{ }\left( 2\left\vert d_{12}\right\vert +2\left\vert
d_{22}\right\vert r_{2}\right) \text{, }2\left\vert d_{13}\right\vert
a_{13}\right\} .  \tag{6.9}
\end{equation}

So, we obtain that $\dot{V}\left( x\right) <0$, when 
\[
x\in \Omega _{0}=\left\{ x\in \Omega _{K},\text{ }\eta _{1}x_{1}+\eta
_{2}x_{2}+\eta _{1}x_{3}<0\right\} . 
\]

Hence, $V\left( x\right) $ is a Lyapunov function\ on the domain $D_{V}$ and
the solution $\bar{x}=\left( \bar{x}_{1},\bar{x}_{2},\bar{x}_{3}\right) $ of
system $\left( 1.2\right) $ satisfying $\left( 6.7\right) $ is global stable
in the sense of Lyapunov.

\begin{center}
\bigskip \textbf{7. Attraction sets for biologically feasible equilibria
points }
\end{center}

\bigskip In this section we will derive global stability of equilibria
points 
\[
\text{ }E_{1}\left( 1,0,0\right) \text{, }E_{2}\left( 0,1,0\right) \text{, }%
E_{3}\left( a,0,b\right) \text{ } 
\]%
and we will find their attraction sets, where 
\[
a=\frac{r_{2}}{a_{21}}\text{, }b=\frac{a_{21}-r_{2}}{a_{13}a_{21}}. 
\]

Let 
\[
\Omega _{K}=\left\{ x\in R^{3}\text{: }0\leq x_{i}\leq K_{i}\text{, }%
i=1,2,3\right\} 
\]%
and 
\[
B_{r}\left( \bar{x}\right) =\left\{ x\in R^{3}\text{, }\left\Vert x-\bar{x}%
\right\Vert _{R^{3}}<r^{2}\right\} . 
\]

\bigskip \textbf{Theorem 7.1. }Assume the following assumptions are
satisfied:

(1) $r_{2}-a_{21}<0,$ $r_{2}+1\leq a_{12};$

(2) $a_{13}>1$, $r_{3}<k_{3}a_{31};$

(3) $d^{2}+\frac{1}{4}\geq 3d+\frac{a_{12}^{2}}{c_{22}^{2}},$ where 
\[
c_{22}=r_{2}+a_{21}-1,\text{ }d=\frac{a_{12}^{2}+c_{22}}{2c_{22}\left(
a_{21}-r_{2}\right) }; 
\]

(4) $\mu =\frac{a_{12}^{2}}{c_{22}}+\left( \frac{a_{12}}{c_{22}}%
-b_{22}\right) r_{2}>0$ and $\frac{d_{3}}{c_{13}}<\min \left\{ 1,\text{ }\mu
\right\} .$

Then the sytem $\left( 1.2\right) $ is global stabile at equilibria point $%
E_{1}\left( 1,0,0\right) $ and the attraction set of the point $E_{1}\left(
1,0,0\right) $ belongs to the set $\Omega _{C}\subset \Omega _{K}\cap \Omega
_{1},$ where

\[
\Omega _{1}=\left\{ x\in \Omega _{K}\text{: }2x_{1}+a_{13}x_{3}<\nu
x_{2}\right\} ,\text{ }\Omega _{C}=\left\{ x\in R^{3}\text{: }V_{1}\left(
x\right) \leq C\text{ }\right\} , 
\]%
here the positive constant $C$ is defined in bellow and 
\[
\nu =\left( a_{12}+\frac{a_{12}}{c_{22}}+\frac{a_{12}}{c_{22}}r_{2}\right) . 
\]

\textbf{Proof. }Let $A_{1}$ be the linearized matrix with respect to
equilibria point $E_{1}\left( 1,0,0\right) ,$ i.e. 
\[
A_{1}=\left[ 
\begin{array}{ccc}
-1 & -a_{12} & 0 \\ 
0 & r_{2}-a_{21} & 0 \\ 
0 & 0 & c_{13}%
\end{array}%
\right] , 
\]%
where 
\[
c_{13}=\frac{r_{3}}{1+k_{3}}-a_{31}-d_{3}<0. 
\]

By assumption (2), $c_{13}<0.$ We consider the Lyapunov equation 
\begin{equation}
P_{1}A_{1}+A_{1}^{T}P_{1}=-I,  \tag{7.1}
\end{equation}%
where 
\[
P_{1}=\left[ 
\begin{array}{ccc}
b_{11} & b_{12} & b_{13} \\ 
b_{21} & b_{22} & b_{23} \\ 
b_{31} & b_{32} & b_{33}%
\end{array}%
\right] . 
\]%
The equation $\left( 7.1\right) $ is reduced to the following linear system
of algebraic equation,%
\[
\begin{array}{c}
2b_{11}=-1, \\ 
-a_{12}b_{11}+\left( r_{2}-a_{21}-1\right) b_{12}=0, \\ 
-\left( c_{13}+1\right) b_{13}=0,%
\end{array}%
\]

\begin{equation}
\begin{array}{c}
-a_{12}b_{11}+\left( r_{2}-a_{21}-1\right) b_{21}=0, \\ 
-a_{12}\left( b_{12}+b_{21}\right) +2\left( r_{2}-a_{21}\right) b_{22}=-1,
\\ 
-a_{21}b_{13}+\left( r_{2}-a_{21}-c_{13}\right) b_{23}=0,%
\end{array}
\tag{7.2}
\end{equation}%
\[
\begin{array}{c}
\left( c_{13}-1\right) b_{31}=0, \\ 
-a_{12}b_{31}+\left( r_{2}-a_{21}-c_{13}\right) b_{32}=0, \\ 
-2c_{13}b_{33}=-1%
\end{array}%
\]

By solving $\left( 7.2\right) $ we obtain 
\[
b_{11}=\frac{1}{2},\text{ }b_{12}=b_{21}=\frac{1}{2}\frac{a_{12}}{c_{22}},%
\text{ }b_{13}=0, 
\]%
\[
\text{ }b_{22}=\frac{a_{12}^{2}+c_{22}}{2\left( a_{21}-r_{2}\right) c_{22}},%
\text{ }b_{23}=0,\text{ }b_{31}=0,\text{ }b_{32}=0,\text{ }b_{33}=\frac{1}{%
2c_{13}}, 
\]

\[
P_{1}=\left[ 
\begin{array}{ccc}
\frac{1}{2} & b_{12} & 0 \\ 
b_{12} & b_{22} & 0 \\ 
0 & 0 & -\frac{1}{2c_{13}}%
\end{array}%
\right] . 
\]

Moreover, 
\begin{equation}
\left\vert P_{1}-\lambda I\right\vert =\left( \lambda -\frac{1}{2c_{13}}%
\right) \left[ \lambda ^{2}-\left( b_{22}+\frac{1}{2}\right) \lambda +\left(
b_{22}+b_{12}^{2}\right) \right] =0.  \tag{7.3}
\end{equation}

From the assumption (2) we deduced $b_{22}>0$. By assumption (3) we have%
\[
\left( b_{22}+\frac{1}{2}\right) ^{2}-4\left( b_{22}+b_{12}^{2}\right) \geq
0. 
\]%
So, $\left( 7.3\right) $ have positive roots, i.e. the matrix $P_{1}$ is
positive defined. Hence, the quadratic function 
\[
V_{1}\left( x\right) =X^{T}P_{1}X=\frac{1}{2}\left( x_{1}-1\right) ^{2}+%
\frac{a_{12}}{c_{22}}\left( x_{1}-1\right) x_{2}+ 
\]%
\[
b_{22}x_{2}^{2}+\frac{1}{2c_{13}}x_{3}^{2} 
\]%
is a positive defined Lyapunov function candidate in certain neighborhood of 
$E_{1}\left( 1,0,0\right) .$ We need now, to determine a domain $\Omega _{1}$
about the point $E_{1},$ where $\dot{V}_{1}\left( x\right) $ is negative
defined and a constant $C$ such that $\Omega _{C}$ is a subset of $\Omega
_{1}$. By assuming $x_{k}\geq 0$, $k=1,2,3,$ we will find the solution set
of the following inequality 
\[
\dot{V}_{1}\left( x\right) =\dsum\limits_{k=1}^{3}\frac{\partial V_{1}}{%
\partial x_{k}}\frac{dx_{k}}{dt}=\left[ \left( x_{1}-1\right) +\frac{a_{12}}{%
c_{22}}x_{2}\right] \left[ \left( 1-x_{1}\right) -a_{12}x_{2}-a_{13}x_{3}%
\right] + 
\]%
\[
\left[ \frac{a_{12}}{c_{22}}\left( x_{1}-1\right) +b_{22}x_{2}\right] x_{2}%
\left[ r_{2}\left( 1-x_{2}\right) -a_{12}x_{1}\right] - 
\]%
\[
-\frac{1}{c_{13}}x_{3}\left[ \frac{r_{3}x_{1}}{x_{1}+k_{3}}-a_{31}x_{1}-d_{3}%
\right] = 
\]%
\[
-\left( 1-x_{1}\right) ^{2}-a_{12}\left( 1-x_{1}\right) x_{2}+a_{13}\left(
1-x_{1}\right) x_{3}-\frac{a_{12}}{c_{22}}\left( 1-x_{1}\right) x_{2}+ 
\]%
\[
-\frac{a_{12}^{2}}{c_{22}}x_{2}^{2}+\frac{a_{12}}{c_{22}}a_{13}x_{2}x_{3}+%
\frac{a_{12}}{c_{22}}r_{2}\left( 1-x_{1}\right) x_{2}\left( 1-x_{2}\right) -%
\frac{a_{12}^{2}}{c_{22}}x_{1}\left( x_{1}-1\right) x_{2}+ 
\]%
\[
b_{22}r_{2}x_{2}^{2}\left( 1-x_{2}\right) -b_{22}a_{12}x_{1}x_{2}^{2}+\frac{1%
}{c_{13}}x_{3}^{2}\left[ \frac{r_{3}x_{1}}{x_{1}+k_{3}}-a_{31}x_{1}-d_{3}%
\right] = 
\]%
\[
-x_{1}^{2}+\left( b_{22}r_{2}-\frac{a_{12}^{2}}{c_{22}}-\frac{a_{12}}{c_{22}}%
r_{2}\right) x_{2}^{2}-\frac{d_{3}}{c_{13}}x_{3}^{2}+\left( a_{12}+\frac{%
a_{12}}{c_{22}}r_{2}+\frac{a_{12}^{2}}{c_{22}}\right) x_{1}x_{2}- 
\]%
\begin{equation}
\frac{a_{12}}{c_{22}}x_{1}x_{2}+\frac{a_{12}}{c_{22}}a_{13}x_{2}x_{3}-\frac{%
a_{12}}{c_{22}}r_{2}x_{1}x_{2}^{2}-\frac{a_{12}^{2}}{c_{22}}x_{1}^{2}x_{2}- 
\tag{7.4}
\end{equation}%
\[
b_{22}r_{2}x_{2}^{3}-b_{22}a_{12}x_{1}x_{2}^{2}+\frac{1}{c_{13}}%
x_{3}^{2}\left( \frac{r_{3}x_{1}}{x_{1}+k_{3}}-a_{31}x_{1}\right) + 
\]%
\[
2x_{1}-\left( a_{12}+\frac{a_{12}}{c_{22}}+\frac{a_{12}}{c_{22}}r_{2}\right)
x_{2}+a_{13}x_{3}-1<0. 
\]

Since for $x\in \Omega _{K},$ 
\[
-\left[ \frac{a_{12}}{c_{22}}r_{2}x_{1}x_{2}^{2}+\frac{a_{12}^{2}}{c_{22}}%
x_{1}^{2}x_{2}+b_{22}r_{2}x_{2}^{3}+b_{22}a_{12}x_{1}x_{2}^{2}\right] \leq
0, 
\]%
\[
-\frac{a_{12}}{c_{22}}x_{1}x_{2}\leq 0,\text{ }-1<0, 
\]%
in view of inequalities 
\begin{equation}
2ab\leq a^{2}+b^{2},\text{ }x_{1}^{2}+x_{2}^{2}\leq \left\Vert x\right\Vert
^{2},\text{ }x_{2}^{2}+x_{3}^{2}\leq \left\Vert x\right\Vert ^{2}  \tag{7.5}
\end{equation}%
the inequality $\left( 7.4\right) $ holds if 
\[
-\left[ x_{1}^{2}+\left( \frac{a_{12}^{2}}{c_{22}}+\frac{a_{12}}{c_{22}}%
r_{2}-b_{22}r_{2}\right) x_{2}^{2}+\frac{d_{3}}{c_{13}}x_{3}^{2}\right] + 
\]%
\begin{equation}
\frac{1}{2}\left( a_{12}+\frac{a_{12}}{c_{22}}r_{2}+\frac{a_{12}^{2}}{c_{22}}%
\right) \left\vert x\right\vert ^{2}+\frac{a_{12}}{2c_{22}}a_{13}\left\vert
x\right\vert ^{2}<0,  \tag{7.6}
\end{equation}%
\[
2x_{1}+a_{13}x_{3}<\left( a_{12}+\frac{a_{12}}{c_{22}}+\frac{a_{12}}{c_{22}}%
r_{2}\right) x_{2}+1, 
\]%
\[
\frac{r_{3}x_{1}}{x_{1}+k_{3}}-a_{31}x_{1}<0. 
\]

\bigskip\ In view of the assumption (1) we get that the solution set of
third inequality of $\left( 7.6\right) $ is $\Omega _{K}.$ By assumption
(4), it is not hard to see that the first inequality $\left( 6.6\right) $ is
satisfied for all $x\in \Omega _{K}$. The solution set of second inequality
in $\left( 7.6\right) $ is the set $\Omega _{1}$. Hence, $\dot{V}_{1}$ is
negative defined on the domain. Hence, $\dot{V}_{1}$ is negative defined on
the domain 
\[
\Omega _{1}=\left\{ x\in B_{r}\left( \bar{x}\right) \text{, }%
2x_{1}+a_{13}x_{3}<\nu x_{2}\right\} =B_{r}\left( \bar{x}\right) \cap \Omega
_{a}, 
\]%
where%
\[
\bar{x}=\left( 1,0,0\right) ,\text{ }r\leq \sqrt{%
K_{1}^{2}+K_{2}^{2}+K_{3}^{2}},\text{ }\nu =a_{12}+\frac{a_{12}}{c_{22}}+%
\frac{a_{12}}{c_{22}}r_{2}, 
\]%
\[
\Omega _{a}=\left\{ x\in \Omega _{K}\text{, }2x_{1}+a_{13}x_{3}<\nu x_{2}%
\text{ }\right\} , 
\]%
i.e., the sytem $\left( 1.2\right) $ is global stabile at $E_{1}\left(
1,0,0\right) $. Let now find the set $\Omega _{C}\subset B_{r}\left( \bar{x}%
\right) ,$ when 
\[
C<\min_{\left\vert x-\bar{x}\right\vert =r}V_{1}\left( x\right) =\lambda
_{\min }\left( P_{1}\right) r^{2}. 
\]%
$\lambda _{\min }\left( P_{1}\right) $ denotes a minimum eigne value of $%
P_{1}$, i.e. 
\[
\lambda _{\min }\left( P_{1}\right) =\min \left\{ \frac{1}{2c_{13}},\text{ }%
\frac{\left( b_{22}+\frac{1}{2}\right) \pm \sqrt{b_{22}^{2}+\frac{1}{4}%
-3b_{22}-4b_{12}^{2}}}{2}\right\} 
\]%
and 
\[
r\leq \sqrt{K_{1}^{2}+K_{2}^{2}+K_{3}^{2}.} 
\]

Moreover, for some $C>0$ the inclusion$\ \Omega _{C}\subset \Omega _{a}$
means the existence of $C>0$ so that $x\in \Omega _{C}$ implies $x\in \Omega
_{a}$, i.e. 
\[
0\leq x_{i}\leq K_{i}\text{, }2x_{1}+a_{13}x_{3}<\nu x_{2}. 
\]

So, 
\[
x\in B_{\bar{r}}\left( \bar{x}\right) =\left\{ x\in R^{3}\text{, }\left\vert
x-\bar{x}\right\vert <\bar{r}\right\} ,
\]%
where 
\[
\text{ }\tilde{r}=\left[ \nu ^{2}\min \left\{ 4\text{, }a_{13}^{2}\right\} +1%
\right] K_{2}^{2}.
\]

Then we obtain that 
\[
C<\min_{\left\vert x\right\vert =r_{1}}V_{1}\left( x\right) =\lambda _{\min
}\left( P_{1}\right) \tilde{r}^{2},
\]%
i.e. 
\[
C<\lambda _{\min }\left( P_{1}\right) r_{0}^{2},\text{ }r_{0}=\min \left\{ r,%
\text{ }\tilde{r}\right\} .
\]

Now, we consider the equilibria point $E_{2}\left( 0,1,0\right) $ and prove
the following result

\textbf{Theorem 7.2. }Assume the following assumptions are satisfied:

(1) $r_{3}<k_{3}a_{31};$

(2) $c_{11}=a_{12}-1>0;$

(3) $\frac{a_{21}^{2}}{c_{11}^{2}}\left( d+1\right) ^{2}+d^{2}\geq \frac{%
a_{21}}{r_{2}c_{11}}\left( d+1\right) ;$

(4) $2b_{12}\left( r_{2}-a_{12}-\frac{1}{2}\left( 1+a_{13}\right) \right)
<\mu ,$ where 
\[
d=\frac{a_{21}}{r_{2}\left( c_{11}+r_{2}\right) },\text{ }\mu =\min \left\{
2b_{11},\text{ }a_{12}\text{, }d_{3}\right\} . 
\]

Then the sytem $\left( 1.2\right) $ is global stabile at equilibria point $%
E_{1}\left( 0,1,0\right) $ and the attraction set of the point $E_{2}\left(
0,1,0\right) $ belongs to the set $\Omega _{C}\subset \Omega _{K}\cap \Omega
_{2},$ where%
\[
\Omega _{2}=\text{ }\left\{ x\in \Omega _{K}\text{:}\right. \text{ }\left(
2b_{11}+\frac{a_{12}}{r_{2}}+2b_{12}\right) x_{1}+2b_{12}\left(
1+a_{12}\right) x_{2}+ 
\]%
\[
\left. 2b_{12}a_{13}x_{3}\leq 0\right\} . 
\]

\[
\text{ }\Omega _{C}=\left\{ x\in R^{3}\text{: }V_{2}\left( x\right) \leq C%
\text{ }\right\} , 
\]%
here $V_{2}\left( x\right) $ and\ the constant $C$ is defined in bellow.

\textbf{Proof. }Let $A_{2}$ be the linearized matrix with respect to
equilibria point $E_{2}\left( 0,1,0\right) ,$ i.e.%
\[
A_{2}=\left[ 
\begin{array}{ccc}
1-a_{12} & 0 & 0 \\ 
-a_{21} & -r_{2} & 0 \\ 
0 & 0 & -d_{3}%
\end{array}%
\right] . 
\]

Consider the Lyapunov equation 
\begin{equation}
P_{2}A_{2}+A_{2}^{T}P_{2}=-I,  \tag{7.7}
\end{equation}%
where 
\[
P_{2}=\left[ 
\begin{array}{ccc}
b_{11} & b_{12} & b_{13} \\ 
b_{21} & b_{22} & b_{23} \\ 
b_{31} & b_{32} & b_{33}%
\end{array}%
\right] . 
\]%
The equation $\left( 7.7\right) $ is reduced to the following linear system
of algebraic equation%
\[
\begin{array}{c}
2c_{11}b_{11}-2a_{21}\left( b_{12}+b_{21}\right) =-1, \\ 
-\left( c_{11}+r_{2}\right) b_{12}-a_{21}b_{22}=0, \\ 
-\left( c_{11}+d_{3}\right) b_{13}-a_{21}b_{23}=0,%
\end{array}%
\]

\begin{equation}
\begin{array}{c}
-\left( c_{11}+r_{2}\right) b_{21}-a_{21}b_{22}=0, \\ 
-2r_{2}b_{22}=-1,\text{ }-\left( d_{3}+r_{2}\right) b_{23}=0,%
\end{array}
\tag{7.8}
\end{equation}%
\[
\begin{array}{c}
-\left( c_{11}+d_{3}\right) b_{31}-a_{21}b_{32}=0, \\ 
-\left( r_{2}+d_{3}\right) b_{32}=0,\text{ }-2d_{3}b_{33}=-1.%
\end{array}%
\]

By solving $\left( 7.8\right) $ we obtain%
\[
b_{13}=0\text{, }b_{12}=b_{21}=\frac{-a_{21}}{2r_{2}\left(
c_{11}+r_{2}\right) },\text{ }b_{11}=\frac{a_{21}}{c_{11}}\left[ \frac{a_{21}%
}{r_{2}\left( c_{11}+r_{2}\right) }+1\right] , 
\]%
\ 
\[
\text{ }b_{22}=\frac{1}{2r_{2}}\text{, }b_{23}=b_{32}=0,\text{ }b_{33}=\frac{%
1}{2d_{3}},\text{ }b_{31}=0, 
\]%
i.e. 
\[
P_{2}=\left[ 
\begin{array}{ccc}
b_{11} & b_{12} & 0 \\ 
b_{12} & \frac{1}{2r_{2}} & 0 \\ 
0 & 0 & \frac{1}{2d_{3}}%
\end{array}%
\right] . 
\]

Moreover, 
\begin{equation}
\left\vert P_{2}-\lambda I\right\vert =\left( \frac{1}{2d_{3}}-\lambda
\right) \left[ \lambda ^{2}-\left( b_{11}+\frac{1}{2r_{2}}\right) \lambda
+\left( \frac{b_{11}}{2r_{2}}-b_{12}^{2}\right) \right] =0.  \tag{7.9}
\end{equation}

In view of the assumption (1) it is clear to see that $b_{11}>0$. By
assumption (3),%
\[
\left( b_{11}+\frac{1}{2r_{2}}\right) ^{2}-4\left( \frac{b_{11}}{2r_{2}}%
-b_{12}^{2}\right) \geq 0\text{.}
\]%
So, $\left( 7.9\right) $ have positive roots, i.e. the matrix $P_{2}$ is
positive defined for all $x$. Hence, the quadratic function 
\[
V_{2}\left( x\right) =X^{T}P_{2}X=b_{11}x_{1}^{2}+2b_{12}x_{1}\left(
x_{2}-1\right) +\frac{1}{2r_{2}}\left( x_{2}-1\right) ^{2}+\frac{1}{2d_{3}}%
x_{3}^{2}
\]%
is a positive defined Lyapunov function candidate in certain neighborhood of 
$E_{2}\left( 0,1,0\right) .$ We need to determine a domain $\Omega _{2}$
about the point $E_{2},$ where $\dot{V}_{2}\left( x\right) $ is negative
defined and a constant $C$ such that $\Omega _{C}$ is a subset of $\Omega
_{2}$. By assuming $x_{k}\geq 0$, $k=1,2,3,$ we will find the solution set
of the following inequality%
\[
\dot{V}_{2}\left( x\right) =\dsum\limits_{k=1}^{3}\frac{\partial V_{2}}{%
\partial x_{k}}\frac{dx_{k}}{dt}=
\]
\[
\left[ 2b_{11}x_{1}+2b_{12}\left( x_{2}-1\right) \right] \left[ \left(
1-x_{1}\right) -a_{12}x_{2}-a_{13}x_{3}\right] +
\]%
\[
\left[ 2b_{12}x_{1}+\frac{1}{r_{2}}\left( x_{2}-1\right) \right] x_{2}\left[
r_{2}\left( 1-x_{2}\right) -a_{12}x_{1}\right] +
\]%
\[
\frac{1}{d_{3}}x_{3}^{2}\left[ \frac{r_{3}x_{1}}{x_{1}+k_{3}}%
-a_{31}x_{1}-d_{3}\right] =-\left(
2b_{11}x_{1}^{2}+a_{12}x_{2}^{2}+d_{3}x_{3}^{2}\right) -2b_{12}a_{12}\left(
x_{1}^{2}+x_{2}^{2}\right) -
\]%
\[
2b_{11}\left( a_{12}x_{1}x_{2}+a_{13}x_{1}x_{3}\right) -2b_{12}\left(
x_{1}x_{2}+a_{13}x_{2}x_{3}\right) -x_{2}\left( x_{2}-1\right) ^{2}-
\]%
\[
\frac{a_{12}}{r_{2}}x_{1}x_{2}+\left( 2b_{11}+\frac{a_{12}}{r_{2}}%
+2b_{12}\right) x_{1}+2b_{12}\left( 1+a_{12}\right) x_{2}+
\]%
\begin{equation}
2b_{12}a_{13}x_{3}+2b_{12}r_{2}\left( x_{1}x_{2}-x_{1}x_{2}^{2}\right) +%
\frac{1}{d_{3}}x_{3}^{2}\left[ \frac{r_{3}x_{1}}{x_{1}+k_{3}}-a_{31}x_{1}%
\right] <0.  \tag{7.10}
\end{equation}

It is clear to see that for all $x\in \Omega _{K}$ 
\begin{equation}
-2b_{11}\left( a_{12}x_{1}x_{2}+a_{13}x_{1}x_{3}\right) <0\text{, }-\frac{%
a_{12}}{r_{2}}x_{1}x_{2}\text{, }-2b_{12}r_{2}x_{1}x_{2}^{2}<0.  \tag{7.11}
\end{equation}

Moreover, in view of $\left( 7.5\right) $ and $\left( 7.11\right) $ the
inequality $\left( 7.10\right) $ holds if

\[
-\left( 2b_{11}x_{1}^{2}+a_{12}x_{2}^{2}+d_{3}x_{3}^{2}\right) +2b_{12}\left[
r_{2}-\left( a_{12}+\frac{1}{2}\left( 1+a_{13}\right) \right) \right]
\left\Vert x\right\Vert ^{2}<0, 
\]

\begin{equation}
\left( 2b_{11}+\frac{a_{12}}{r_{2}}+2b_{12}\right) x_{1}+2b_{12}\left(
1+a_{12}\right) x_{2}+2b_{12}a_{13}x_{3}\leq x_{2}\left( x_{2}-1\right) ^{2},
\tag{7.12}
\end{equation}

\[
\frac{r_{3}x_{1}}{x_{1}+k_{3}}-a_{31}x_{1}<0. 
\]

\ In view of assumption (1) the solution set of third inequality of $\left(
6.13\right) $ is $\Omega _{K}$. By assumption (4), it is not hard to see
that the first inequality $\left( 7.12\right) $ is satisfied for all $x\in
\Omega _{K}$. The solution set of second inequality in $\left( 7.12\right) $
is the set $\Omega _{2}$. Hence, $\dot{V}_{2}$ is negative defined on the
domain%
\[
\Omega _{0}=B_{r}\left( \bar{x}\right) \cap \Omega _{2}\text{, }r\leq
K_{1}^{2}+K_{2}^{2}+K_{3}^{2}, 
\]%
i.e., the sytem $\left( 1.2\right) $ is global stabile at $E_{1}\left(
0,1,0\right) .$ We will find $C>0$ such that $\Omega _{C}\subset B_{r}\left( 
\bar{x}\right) \cap \Omega _{2}$, i.e. 
\[
\Omega _{C}\subset B_{r}\left( \bar{x}\right) \cap \Omega _{2}. 
\]

\bigskip It is clear to see that $\Omega _{C}\subset B_{r}\left( \bar{x}%
\right) ,$ when 
\[
C<\min_{\left\vert x-\bar{x}\right\vert =r}V_{2}\left( x\right) =\lambda
_{\min }\left( P_{2}\right) r^{2},\text{ }\bar{x}=\left( 0,1,0\right) , 
\]
here $\lambda _{\min }\left( P_{2}\right) $ denotes a minimum eigne value of 
$P_{2}$, i.e. 
\[
\lambda _{\min }\left( P_{2}\right) =\min \left\{ \frac{1}{2d_{3}},\text{ }%
\frac{\left( b_{22}+\frac{1}{2r_{2}}\right) \pm \sqrt{b_{11}^{2}+b_{12}^{2}+%
\frac{1}{4r_{2}^{2}}-\frac{b_{11}}{r_{2}}}}{2}\right\} . 
\]

Moreover, for some $C>0$ the inclusion$\ \Omega _{C}\subset \Omega _{2}$
means the existence of $C>0$ so that $x\in \Omega _{C}$ implies $x\in \Omega
_{2}$, i.e.%
\begin{equation}
x\in \Omega _{K}\text{, }\left( 2b_{11}+\frac{a_{12}}{r_{2}}+2b_{12}\right)
x_{1}+2b_{12}\left( 1+a_{12}\right) x_{2}+2b_{12}a_{13}x_{3}\leq 0. 
\tag{7.13}
\end{equation}

\bigskip $\left( 7.13\right) $ implies 
\[
\left( 2b_{11}+\frac{a_{12}}{r_{2}}\right) x_{1}\leq -2b_{12}\left[ \left(
1+a_{12}\right) x_{2}+a_{13}x_{3}\right] , 
\]

So, 
\[
x\in B_{\bar{r}}\left( \bar{x}\right) =\left\{ x\in R^{3}\text{, }\left\vert
x-\bar{x}\right\vert <\bar{r}\right\} ,
\]%
where 
\[
\eta _{1}=\frac{-2b_{12}\left( 1+a_{12}\right) }{2b_{11}+\frac{a_{12}}{r_{2}}%
}\text{, }\eta _{2}=\frac{-2b_{12}a_{13}}{2b_{11}+\frac{a_{12}}{r_{2}}},%
\text{ }\bar{r}=\left[ 2\left( 1+\eta _{1}^{2}\right) K_{2}^{2}+2\left(
1+\eta _{2}^{2}\right) K_{3}^{2}\right] ^{\frac{1}{2}}.
\]%
Then we obtain that 
\[
C<\min_{\left\vert x\right\vert =r_{0}}V_{2}\left( x\right) =\lambda _{\min
}\left( P_{2}\right) \bar{r}^{2},
\]%
i.e. 
\[
C<\lambda _{\min }\left( P_{2}\right) \bar{r}^{2}\text{ for }r_{0}=\min
\left\{ r,\text{ }\bar{r}\right\} .
\]

\textbf{Remark 7.1. }It is clear to see that if $a_{21}\geq \frac{a_{12}-1}{%
r_{2}}$, then the assumption (3) is satisfied. Moreover, if $a_{12}+\frac{1}{%
2}\left( 1+a_{13}\right) >r_{2},$ then the assumption (4) holds.

Let $A_{3}$ be the linearized matrix with respect to equilibria point $%
E_{3}\left( a,0,b\right) $ defined by $\left( 4.11\right) $, i.e.,%
\[
A_{3}=\left[ 
\begin{array}{ccc}
-a & d_{12} & d_{13} \\ 
0 & d_{22} & 0 \\ 
d_{31} & 0 & d_{33}%
\end{array}%
\right] , 
\]%
where 
\[
d_{12}=-a_{12}a\text{, }d_{13}=-a_{13}a\text{, }d_{22}=r_{2}-a_{21}a,\text{ }%
a=\frac{r_{2}}{a_{21}},\text{ }d_{31}= 
\]%
\[
\left( \frac{k_{1}r_{3}}{\left( a+k_{1}\right) ^{2}}-a_{31}\right) b,\text{ }%
d_{33}=\frac{r_{3}a}{a+k_{3}}-a_{31}a-d_{3}\text{, }b=\frac{a_{21}-r_{2}}{%
a_{13}a_{21}}. 
\]

By reasoning as the above we obtain

\textbf{Theorem 7.3. }Assume the Condition 5.6 hold and:

(1) $a_{31}>\frac{k_{1}r_{3}}{\left( a+k_{1}\right) ^{2}},$ $a_{31}+d_{3}>%
\frac{r_{3}a}{\left( a+k_{1}\right) ^{2}};$

(2) 
\[
\frac{a_{31}-\frac{k_{1}r_{3}}{\left( a+k_{1}\right) ^{2}}}{d_{3}+a_{31}-%
\frac{r_{3}a}{\left( a+k_{3}\right) }}>\frac{1}{a\left( a_{12}+a_{13}\right) 
}. 
\]

Then the sytem $\left( 1.2\right) $ is global stabile at equilibria point $%
E_{3}\left( a,0,b\right) $ and the attraction set of the point $E_{3}\left(
a,0,b\right) $ belongs to the set $\Omega _{C}\subset \Omega _{K}\cap \Omega
_{3},$ where

\[
\Omega _{3}=\left\{ x=\left( x_{1},x_{2},x_{3}\right) \in \Omega
_{K}:\right. 
\]

\begin{equation}
\alpha x_{1}+\alpha _{2}x_{2}+\alpha _{3}x_{3}\geq 0,\text{ }\alpha
x_{1}+\beta _{2}x_{2}+\left. \beta _{3}x_{3}\geq 0\right\} ,  \tag{7.14}
\end{equation}

\[
\text{ }\Omega _{C}=\left\{ x\in R^{3}\text{: }V_{3}\left( x\right) \leq C%
\text{ }\right\} ,
\]%
here $V_{3}\left( x\right) $ and\ the constant $C$ is defined in bellow and 
\[
\alpha =\frac{d_{12}}{\left( \mu a+d_{12}d_{31}\right) }\text{, }\alpha _{2}=%
\frac{2\left( 1+d_{12}b_{12}\right) }{d_{12}+2d_{22}},\text{ }\alpha _{3}=%
\frac{2\left( d_{12}+d_{13}\right) }{d_{22}+d_{33}}b_{12},\text{ }
\]%
\[
\beta _{2}=\frac{2\left( d_{12}+d_{13}\right) }{d_{22}+d_{33}}b_{12}\text{, }%
\beta _{3}=\frac{1}{d_{33}}\left( 1+2d_{13}b_{12}\right) .
\]%
.

\textbf{Proof. }Consider the Lyapunov equation 
\begin{equation}
P_{3}A_{3}+A_{3}^{T}P_{3}=-I,  \tag{7.15}
\end{equation}%
where 
\[
P_{3}=\left[ 
\begin{array}{ccc}
b_{11} & b_{12} & b_{13} \\ 
b_{21} & b_{22} & b_{23} \\ 
b_{31} & b_{32} & b_{33}%
\end{array}%
\right] \text{, }b_{ij}=b_{ji}. 
\]

\bigskip By solv\i ng algebraic equation in $b_{ij}$ according to $\left(
7.15\right) $ we obtain 
\[
\text{ }b_{12}=b_{13}=-\frac{d_{12}}{2\left( \mu a+d_{12}d_{31}\right) },%
\text{ }b_{11}=-\frac{\mu b_{12}}{d_{12}},\text{\ }b_{22}=-\frac{\left(
1+d_{12}b_{12}\right) }{d_{12}+2d_{22}}, 
\]%
\begin{equation}
\text{ }b_{23}=-\frac{d_{12}+d_{13}}{d_{22}+d_{33}}b_{12},\text{ }b_{33}=-%
\frac{1}{2d_{33}}\left( 1+2d_{13}b_{12}\right) ,\text{ }  \tag{7.16}
\end{equation}%
\[
\mu =d_{22}-a-\frac{\left( d_{12}+d_{13}\right) d_{31}}{d_{22}+d_{33}}. 
\]%
Then by using the assumptions (1) and (2) we get that that $P_{3}$ have
positive eigne values, i.e. the quadratic function 
\[
V_{3}\left( x\right) =X^{T}P_{3}X=-\frac{\mu }{d_{12}}x_{1}^{2}-\frac{\left(
1+d_{12}b_{12}\right) }{d_{12}+2d_{22}}x_{2}^{2}-\frac{1}{2d_{33}}\left(
1+2d_{13}b_{12}\right) x_{3}^{2}- 
\]%
\[
-\frac{d_{12}}{\left( \mu a+d_{12}d_{31}\right) }\left(
x_{1}x_{2}+x_{1}x_{3}\right) -\frac{2\left( d_{12}+d_{13}\right) }{%
d_{22}+d_{33}}b_{12}x_{2}x_{3}. 
\]%
is a positive defined Lyapunov function candidate in neighborhood of $%
E_{3}\left( a,0,b\right) .$

By assuming $x_{k}\geq 0$, $k=1,2,3,$ we find that 
\begin{equation}
\dot{V}_{3}\left( x\right) =\dsum\limits_{k=1}^{3}\frac{\partial V_{3}}{%
\partial x_{k}}\frac{dx_{k}}{dt}=  \tag{7.17}
\end{equation}%
\[
-\left[ \frac{2\mu }{d_{12}}x_{1}+\frac{d_{12}}{\left( \mu
a+d_{12}d_{31}\right) }\left( x_{2}+x_{3}\right) \right] x_{1}\left[ \left(
1-x_{1}\right) -a_{12}x_{2}-a_{13}x_{3}\right] -
\]%
\[
\left( \alpha x_{1}+\alpha _{2}x_{2}+\alpha _{3}x_{3}\right) \left[
r_{2}\left( 1-x_{2}\right) -a_{12}x_{1}\right] -
\]%
\[
\left( \alpha x_{1}+\beta _{2}x_{2}+\beta _{2}x_{2}\right) \left[ \frac{%
r_{3}x_{1}}{x_{1}+k_{3}}-a_{31}x_{1}-d_{3}\right] =
\]%
\[
\frac{2\mu }{d_{12}}\left[ -x_{1}^{2}\left( 1-x_{1}\right)
+a_{12}x_{1}x_{2}+a_{13}x_{1}x_{3}\right] -\alpha \left(
x_{1}x_{2}+x_{1}x_{3}\right) \left( 1-x_{1}\right) +
\]%
\[
\alpha \left[ a_{12}\left( x_{2}^{2}+x_{2}x_{3}\right) +a_{13}\left(
x_{3}^{2}+x_{2}x_{3}\right) \right] -\frac{2\left( 1+d_{12}b_{12}\right)
r_{2}}{d_{12}+2d_{22}}x_{2}\left( 1-x_{2}\right) +
\]%
\[
\frac{2\left( 1+d_{12}b_{12}\right) a_{12}}{d_{12}+2d_{22}}x_{1}x_{2}-\alpha
r_{2}x_{1}\left( 1-x_{2}\right) +\frac{d_{12}r_{2}a_{12}}{\left( \mu
a+d_{12}d_{31}\right) }x_{1}^{2}+
\]%
\[
-\frac{2\left( d_{12}+d_{13}\right) r_{2}}{d_{22}+d_{33}}b_{12}x_{3}\left(
1-x_{2}\right) +\frac{2\left( d_{12}+d_{13}\right) a_{12}}{d_{22}+d_{33}}%
b_{12}x_{1}x_{3}-
\]%
\[
\left( \alpha x_{1}+\beta _{2}x_{2}+\beta _{2}x_{2}\right) \left[ \frac{%
r_{3}x_{1}}{x_{1}+k_{3}}-a_{31}x_{1}-d_{3}\right] .
\]

By assumption (1), (2), $\mu >0,$ $d_{12}<0,$ $d_{13}<0$. So, for $x\in
\Omega _{K}$ we have 
\[
\frac{2\mu }{d_{12}}x_{1}^{3}\leq 0\text{, }\frac{d_{12}}{\left( \mu
a+d_{12}d_{31}\right) }\left( x_{1}x_{2}+x_{1}x_{3}\right) x_{1}\leq 0.\text{
} 
\]%
In view of $\left( 7.14\right) $ the estimate$\ \left( 7.17\right) $ holds
if the following inequalitie is satisfied

\[
\frac{2\mu }{d_{12}}\left[ -x_{1}^{2}+a_{12}x_{1}x_{2}+a_{13}x_{1}x_{3}%
\right] -\frac{d_{12}}{\left( \mu a+d_{12}d_{31}\right) }\left(
x_{1}x_{2}+x_{1}x_{3}\right) + 
\]%
\[
\frac{d_{12}}{\left( \mu a+d_{12}d_{31}\right) }\left[ a_{12}\left(
x_{2}^{2}+x_{2}x_{3}\right) +a_{13}\left( x_{3}^{2}+x_{2}x_{3}\right) \right]
+\frac{2\left( 1+d_{12}b_{12}\right) r_{2}}{d_{12}+2d_{22}}x_{2}^{2}+ 
\]%
\[
\left[ \frac{2\left( 1+d_{12}b_{12}\right) a_{12}}{d_{12}+2d_{22}}+\frac{%
d_{12}r_{2}}{\left( \mu a+d_{12}d_{31}\right) }\right] x_{1}x_{2}+\frac{%
d_{12}r_{2}a_{12}}{\left( \mu a+d_{12}d_{31}\right) }x_{1}^{2}+ 
\]%
\begin{equation}
\frac{2\left( d_{12}+d_{13}\right) r_{2}}{d_{22}+d_{33}}b_{12}x_{2}x_{3}+%
\frac{2\left( d_{12}+d_{13}\right) a_{12}}{d_{22}+d_{33}}b_{12}x_{1}x_{3}<0.
\tag{7.18}
\end{equation}

By Condition 5.6 and by assumptions (1), (2) the inequality $\left(
7.18\right) $ is satisfied for $x\in \Omega _{K}.$ So, $\left( 7.17\right) $
holds if the inequalities $\left( 7.14\right) $ and $\left( 7.18\right) $
are satisfied for $x\in \Omega _{K}.$ Hence, $\dot{V}_{3}$ is negative
defined on the domain%
\[
\Omega _{0}=B_{r}\left( \bar{x}\right) \cap \Omega _{3}\text{, }r\leq
K_{1}^{2}+K_{2}^{2}+K_{3}^{2},
\]%
i.e., the sytem $\left( 1.2\right) $ is global stabile at the point $%
E_{3}\left( a,1,b\right) .$ We will find $C>0$ such that $\Omega _{C}\subset
B_{r}\left( \bar{x}\right) \cap \Omega _{3}$, i.e. 
\[
\Omega _{C}\subset B_{r}\left( \bar{x}\right) \cap \Omega _{3}.
\]

\bigskip It is clear to see that $\Omega _{C}\subset B_{r}\left( \bar{x}%
\right) ,$ when 
\[
C<\min_{\left\vert x-\bar{x}\right\vert =r}V_{2}\left( x\right) =\lambda
_{\min }\left( P_{2}\right) r^{2},\text{ }\bar{x}=\left( a,0,b\right) ,
\]%
here $\lambda _{\min }\left( P_{3}\right) $ denotes a minimum eigne value of 
$P_{3}$.

Moreover, for some $C>0$ the inclusion$\ \Omega _{C}\subset \Omega _{3}$
means the existence of $C>0$ so that $x\in \Omega _{C}$ implies $x\in \Omega
_{3}$, i.e. $\left( 7.17\right) $ holds.

So, 
\[
x\in B_{\bar{r}}\left( \bar{x}\right) =\left\{ x\in R^{3}\text{, }\left\vert
x-\bar{x}\right\vert <\bar{r}\right\} ,
\]%
where 
\[
\mu _{2}=\min \left\{ -\frac{\alpha _{2}}{\alpha }\text{, }-\frac{\beta _{2}%
}{\alpha }\right\} \text{, }\mu _{3}=\left\{ -\frac{\alpha _{3}}{\alpha }%
\text{, }-\frac{\beta _{3}}{\alpha }\right\} ,\text{ }
\]%
\[
\bar{r}=\left[ 2\left( 1+\mu _{2}^{2}\right) K_{2}^{2}+2\left( 1+\mu
_{3}^{2}\right) K_{3}^{2}\right] ^{\frac{1}{2}}.
\]%
Then we obtain that 
\[
C<\min_{\left\vert x\right\vert =r_{0}}V_{3}\left( x\right) =\lambda _{\min
}\left( P_{3}\right) \bar{r}^{2},
\]%
i.e. 
\[
C<\lambda _{\min }\left( P_{3}\right) \bar{r}^{2}\text{ for }r_{0}=\min
\left\{ r,\text{ }\bar{r}\right\} .
\]

\begin{center}
\textbf{Acknowledgements}
\end{center}

The author is thanking to Assist. Prof. of department of biophysics,
Yeditepe University A. Maharramov, Assoc. Prof. of department of Immunology
Yeditepe University G\"{u}lderen Yan\i kkaya Demirel and Prof. Dr. Faculty
of Helth Sciences of Okan University Aida Sahmurova according to their
valuable suggestions in the field of medicine and biology.

\textbf{References}

\begin{enumerate}
\item Kuznetsov V. A., Makalkin I. A., Taylor M. A., Perelson A. S.,
Nonlinear dynamics of immunogenic tumors: parameter estimation and global
bifurcation analysis, Bull. Math. Biol. 1994(56), 295--321.

\item Adam J. A, Bellomo C., A survey of models for tumor-immune system
dynamics, Boston, MA: Birkhauser, 1996.

\item Eftimie R, Bramson J. L., Earn D. J. D., Interactions between the
immune system and cancer: a brief review of non-spatial mathematical models,
Bull. Math. Biol. 2011(73), 2--32.

\item Kirschner D, Panetta J., Modelling immunotherapy of the tumor--immune
interaction, J. Math. Biol. 1998(37), 235--52.

\item de Pillis L. G., Radunskaya A., The dynamics of an optimally
controlled tumor model: a case study, Math. Comput. Modell 2003(37),1221--44.

\item Nani F., Freedman H. A., Mathematical model of cancer treatment by
immunotherapy, Math Biosci. 2000(163), 159--99.

\item Altrock P. M., Lin L. Liu., Michor F., The mathematics of cancer:
integrating quantitative models, Nature Reviews Cancer. 15 (2015), No.12,
730-745.

\item Itik I. M., Banks S. P., Chaos in a three-dimensional cancer model,
Int J. Bifurcation Chaos 2010, 2010(20), 71--79.

\item Levine, H., A., Pamuk S., Sleeman B. D., Mathematical modeling of
capillary formation and development in tumor angiogenesis, Penetration into
the Stroma bulletin of Mathematical Biology (2001) 63, 801--863.

\item Starkov, K. E., Krishchenko A. P., On the global dynamics of one
cancer tumour growth model, Commun. Nonlinear Sci. Numer. Simul. 19 (2014),
1486--1495.

\item Khalil H. Nonlinear systems, NJ: Prentice Hall, 2002.
\end{enumerate}

\end{document}